\documentclass[iop]{emulateapj}
\usepackage{graphicx}
\usepackage[flushleft]{threeparttable}
\usepackage{blindtext}
\usepackage{amsmath}
\usepackage{mathtools}
\usepackage{multirow}
\usepackage{hyperref}
\hypersetup{
	colorlinks	= true,
	linkcolor	= red,
	urlcolor	= cyan,
	citecolor	= blue
}
\usepackage[usenames, dvipsnames]{color}

\newcommand{\stsci}{\mbox{STScI}}
\newcommand{\mast}{\mbox{MAST} Archive}
\newcommand{\hst}{\mbox{HST}}
\newcommand{\wfc}{\mbox{WFC3}}
\newcommand{\stis}{\mbox{STIS}}
\newcommand{\acs}{\mbox{ACS}}
\newcommand{\coss}{\mbox{COS}}
\newcommand{\grisml}{\mbox{G141}}
\newcommand{\grisms}{\mbox{G102}}

\newcommand{\axe}{\mbox{aXe}}
\newcommand{\wfcsim}{\mbox{\textit{Wayne}}}
\newcommand{\axesim}{\mbox{aXeSim}}
\newcommand{\pylc}{\mbox{PyLightcurve}}

\newcommand{\system}{\mbox{HD\,209458}}
\newcommand{\planet}{\mbox{HD\,209458\,b}}

\begin{document}
	
\title{Wayne - A Simulator for HST WFC3 IR Grism Spectroscopy}

\author{R. Varley\altaffilmark{1}, A. Tsiaras\altaffilmark{1} and K. Karpouzas\altaffilmark{2}}

\affil{$^1$Department of Physics \& Astronomy, University College London, Gower Street, WC1E6BT London, United Kingdom}
\affil{$^2$Department of Physics, Section of Astrophysics, Astronomy and Mechanics, \\ Aristotle University of Thessaloniki, 541 24 Thessaloniki, Greece}

\email{r.varley@ucl.ac.uk}

\begin{abstract}
\wfcsim\ is an algorithm that simulates Hubble Space Telescope (\hst) Wide Field Camera 3 (\wfc) grism spectroscopic frames including sources of noise and systematics. It can simulate both staring and spatial scan modes, and observations such as the transit and the eclipse of an exoplanet.
Unlike many other instrument simulators, the focus of \wfcsim\ is on creating frames with realistic systematics in order to test the effectiveness of different data analysis methods in a variety of different scenarios. This approach is critical for method validation and optimising observing strategies.
In this paper we describe the implementation of \wfcsim\ for \wfc\ in the near-infrared channel with the \grisms\ and \grisml\ grisms. We compare the simulations to real data, obtained for the exoplanet \planet\ to verify the accuracy of the simulation. The software is now available as open source at \url{https://github.com/ucl-exoplanets/wayne}.

\end{abstract}


\maketitle

\section{Introduction}

The Hubble Space Telescope (\hst) has been invaluable for exoplanetary science. The \stis\ spectrograph was first used for time-series photometry of \planet\ \citep{Brown2001} and then for the first measurement of an exoplanet atmosphere with transmission spectroscopy \citep{Charbonneau2002}. It has enabled several atomic and ionic detections \citep[e.g.][]{VidalMadjar2003, VidalMadjar2004} as well as scattering properties \citep[e.g.][]{Sing2011, Sing2015}. The \acs\ and \coss\ instruments have both made measurements of exoplanet atmospheres \citep[e.g][]{Pont2008, Linksky2010}. NICMOS was used to search for molecules in an exoplanet atmosphere \citep[e.g.][]{Swain2008, Tinetti2010, Crouzet2012} and has directly imaged exoplanets \citep[e.g.][]{Song2006}. In 2009 the Wide Field Camera 3 (\wfc) instrument was installed and has provided many significant measurements in the near-infrared (IR) channel \citep[e.g.][]{Berta2012, Swain-wasp12, Kreidberg-gj1214b}. Whilst \wfc\ has been successfully used, as with any instrument, it presents some unique challenges for data analysis. This is particularly true when used in spatial scan mode for exoplanet transit and eclipse spectroscopy, which is the focus of our paper.

\subsection{Staring Mode}
Exoplanet transit spectra were initially observed with \hst/\wfc\ using the traditional ``staring'' mode, where the telescope pointing is fixed on the target star \citep[e.g][]{Berta2012, Swain-wasp12, Wilkins2014}. The main systematic found in these data is the ``ramp'' or ``hook'' effect which causes a sharp rise in flux before quickly plateauing after each \hst\ buffer dump. The other important effect is a gradual reduction in the level of flux throughout the observation. The analysis by \citet{Berta2012} found the amplitude of the ``hook'' to be around 0.4\% and the visit long slope to be around 0.05\%. These are both much larger than the 10$^{-4}$ variations expected in the planetary spectrum. In addition, the initial stage of the analysis includes reductions such as flat-fielding and a non-linearity correction which, despite not being expected to cause issues in the analysis, their end-to-end effect and coupling with other noise sources and systematics has not been fully investigated yet.

Data taken by \hst\ is also non continuous as \hst\ is in a low earth orbit. A typical transit event observed by \hst\ lasts around one to four hours, meaning an observation must span multiple orbits, as a single target is only continuously visible for $\sim$\,45 minutes. This time includes overheads such as acquiring and re-acquiring the guide star \citep[6 and 5 minutes respectively,][pg. 209]{WFC3InstHandbook} and buffer dumps. The buffer is a temporary storage space for images on \hst\ which must be transferred (dumped) to the solid state recorder when full. A buffer dump is required after the instrument has taken a certain number of exposures and takes $\sim$\,6 minutes \citep[pg. 212]{WFC3InstHandbook} during which no new exposures are taken. Buffer dumps are generally required when taking many short exposures (usually one or two dumps per orbit for staring data), and lower the overall efficiency of an observation.

\subsection{Spatial Scan Mode}
In 2011 (\hst\ Cycle 19) the spatial scan mode was introduced. This mode slews the telescope during an exposure under the control of either the fine-guidance sensors (FGS, for scans speeds below 5$''$\,s$^{-1}$) or gyroscope only control (for scan speeds up to 7$''$.84\,s$^{-1}$). The advantages of this mode over staring include the ability to observe targets that would normally saturate the detector and to achieve higher signal-to-noise ratios (S/N), due to reduced overheads (as each individual exposure can last much longer before saturation). These overheads include the reduced number of detector reads and the reduction, or elimination, of mid-orbit buffer dumps. Many observations have been performed in spatial scan mode \citep[e.g][]{Deming2013, Knutson2014a, Knutson2014b, Kreidberg-gj1214b, Kreidberg-wasp43b, Kreidberg-wasp12b, McCullough2014, Crouzet2014, Fraine2014, Stevenson2014} however there are some caveats. In addition to the hook present in staring mode, the use of the FGS to control the scan introduces some jitter, causing a periodic ``flicker'' in the brightness along the scan direction due to FGS feedback \citep{WFC3ScancCalib2012}. Finally, in spatial scan mode there are dispersion variations along the scanning direction \citep{Tsiaras2015}, shifts in the x and y position of the spectrum in each exposure  \citep{Deming2013} and changes in the effective exposure time when scanning in different directions \citep[the upstream / downstream effect, see][]{WFC3ScancCalib2012, Knutson2014a, Tsiaras2016}.

Since instrumental systematics are of the order of the planetary signal, and as observations typically consist of a single transit, it is hard to verify beyond any reasonable doubt whether the result obtained is in fact the independent planetary signal, not coupled with systematics or residuals from the analysis. This has led to approaches using machine learning algorithms \citep[e.g.][]{Waldmann2013, Morello2014} which attempt to extract the signal `unsupervised'. 

In an ideal world we would calibrate our instruments and pipelines with a host of known stable reference stars. In absence of this, we propose here a simulator capable of replicating the measured data, including systematics. Such a tool will help us explore the effects of different instrument systematics and observing techniques on the final scientific result. Moreover, we will be able to understand the coupling between those effects, a study that cannot be carried out based only on real observations, for which the input signal is not known. We can then test the ability of pipelines and reduction methods to recover the planetary signal in different scenarios, which is an important verification step in data analysis. We can, also, validate the stability and reliability of existing data sets, by taking into account the specific configuration that each one has. Finally, such a simulator can be used to explore different possible configurations for future observations and evaluate their effect on the retrieved planetary spectrum.

To this end, we describe here the implementation of \wfcsim, a simulator for \wfc\ IR spectroscopy in both the \grisms\ and \grisml\ grisms. We implemented many systematics and noise sources present in real data and \wfcsim\ is able to simulate both ordinary stellar spectra as well as exoplanet transit spectroscopy, where the planetary and stellar spectra are combined together in a transit event. We demonstrate the accuracy of our simulation by replicating the real observation of \planet\ from \hst\ proposal 12181 and process it using the same pipeline described in \citet{Tsiaras2015}. We compare the real observation to our simulation providing additional verification of the analysis for that observation.

\newpage

\subsection{Other Simulators}

Comparisons are obviously drawn to \axesim \footnote{\url{http://axe.stsci.edu/axesim/}}, the \stsci\ grism simulator package for \wfc. \axesim\ was designed for general astrophysics whereas \wfcsim\ has been conceived for transit and eclipse spectroscopy. The limitations of \axesim, which are accounted for in \wfcsim, include spatial scan mode, scan speed variations, dispersion variations, x and y positional shifts, trends such as the ``hook'' and long term ramp, cosmic rays and the ability to simulate a transit with time. However, \wfcsim\ is currently only able to simulate a single spectrum at a time whereas \axesim\ can simulate a whole field at once. We also do not attempt to reimplement the functionality of the Hubble Space Telescope Exposure Time Calculator\footnote{\url{http://www.stsci.edu/hst/wfc3/tools/etcs/}}, which is able to approximate the signal to noise of a target. We instead rely on the user to scale the input flux of the star to the correct level using the exposure calculator or otherwise. The software is now available as open source at \url{https://github.com/ucl-exoplanets/wayne}.

\section{\wfcsim}

The simulator imitates real data taken by \wfc\ IR grisms through four key steps:

\begin{itemize}
\item \textbf{Preparing the visit} --- Calculation of the exposure time of each frame including any gaps during orbits and buffer dumps.
\item \textbf{Preparing the spectrum} --- Scaling the flux to the instrument, applying the grism sensitivity, any re-binning and, in the case of transit spectroscopy, combining the stellar and planetary signals.
\item \textbf{Converting flux into pixels} --- Calculating where on the detector a specific wavelength falls, simulating the point-spread function and integrating it into pixels.
\item \textbf{Adding any noise and implementing reductions} --- Applying the flat field, scan speed variations, detector gain, read noise etc.
\end{itemize}

\begin{figure}
        \centering
        \includegraphics[width=\columnwidth]{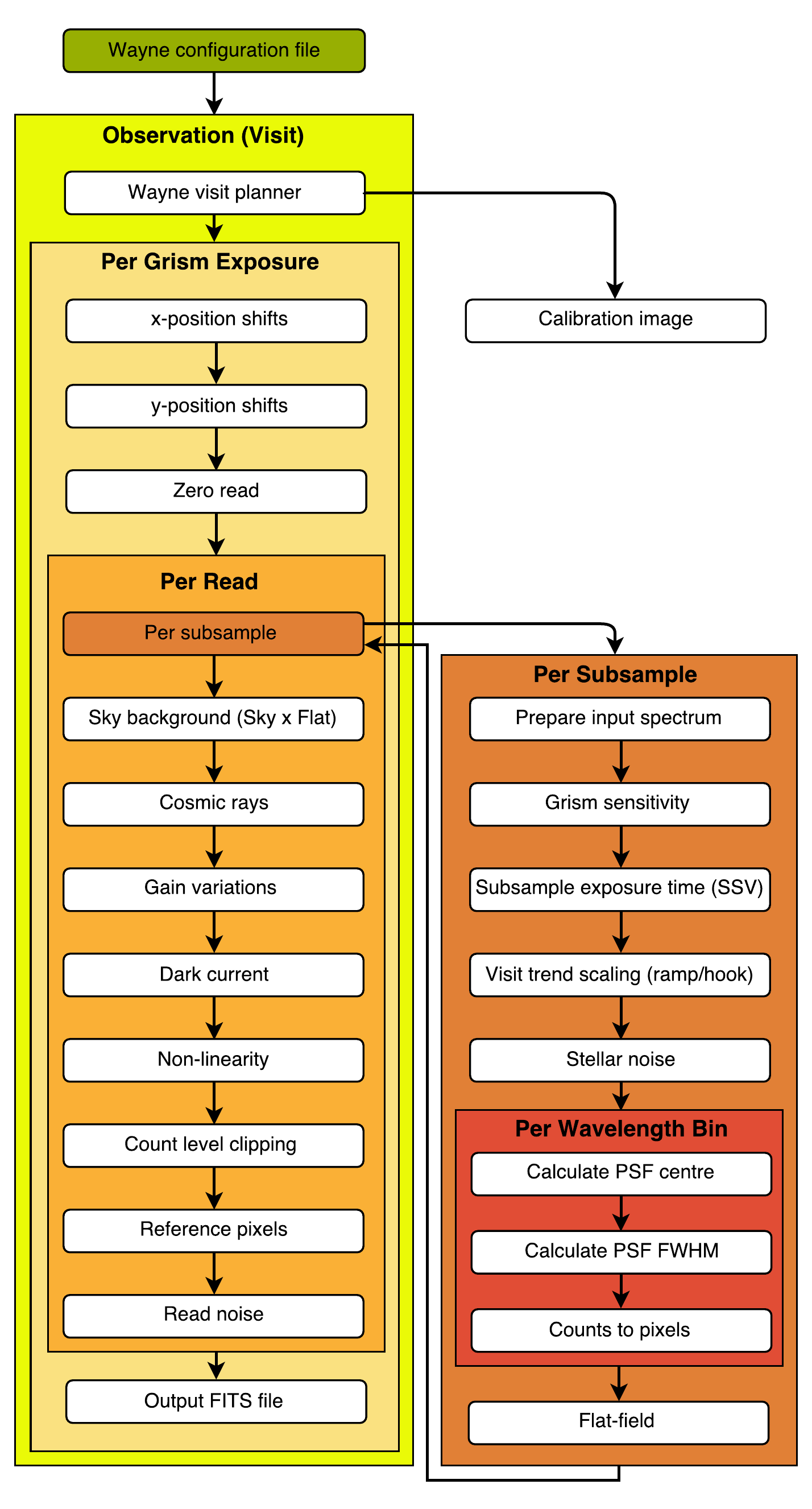}
        \caption{Simplified overview of the steps performed by \wfcsim\ in generating a visit. Steps are shown where they are applied, while the actual calculation may be performed at a different location for computational optimisation reasons.}
        \label{fig:exposure-overview-flow}
\end{figure}

These steps are often overlapping as reductions and systematics need to be applied at different stages of the simulation. A simplified overview of the process used to create a simulation is shown in Figure \ref{fig:exposure-overview-flow}.

There are many sources of noise and trends caused by the telescope and its instruments. Some of these, such as the dark current, flat-field and quantum efficiency have been constrained very well through calibration programs and are used in reduction pipelines like \axe\ \citep{axe2009}. Others, like the variation in scan speed and the hook trend are not yet well understood or calibrated. For the first set we used the state of the art from the literature. For the latter, we conducted our own investigations on real data sets to determine approximate models for these features, and allow their parameters in the simulations to be varied by the user. We plan to investigate these behaviours in the future.

The \wfc\ detector is read multiple times per exposure in what is known as a non-destructive read or ``sample up the ramp''. We refer to these as ``reads''. In order to generate a spatial scan exposure we must combine many staring mode frames together at intervals in time (and therefore position). We use the term ``subsample'' to refer to these to avoid confusion. The exposure time of a subsample is set in the configuration file of the simulator and is generally 5--30\,ms. The exact, i.e computationally optimised, value of this parameter depends on the scan speed.

In this section we mention certain files such as the `flat-field cube' which contains the corrections we are using. These are mostly obtained from \stsci\ and we provide the location for each file in the appendix \ref{sec:correction-files}. 

In this paper we use the transit spectroscopy mode while spatial scanning as an example of \wfcsim. The \grisms\ and \grisml\ grism implementations are very similar, differing only in the calibration files used. As such, we focus on the \grisml\ grism as it is more commonly used in exoplanet transit spectroscopy.

\subsection{Planning a Visit}
\hst\ is located in a low earth orbit, completing an orbit of Earth in around 95 minutes. Single targets are typically visible for 50 minutes of each orbit but can be longer depending on their location with respect to \hst\ orbit. Visits comprise of multiple orbits, typically limited to 5 or 6 due to the South Atlantic Anomaly \citep[pg. 17]{WFC3Primer23}. The first orbit in a visit exhibits the most severe trends due to spacecraft settling and is therefore normally left out of the analysis. For this reason, we do not attempt to recreate the first orbit effects in our simulations. 

\wfc\ IR detector has several different observing modes which set the exposure time and the size of the detector array that is read. These parameters are defined by three variables, the subarray mode (\textit{SUBARRAY}=1024, 512, 256, 128, 64), the sample sequence (\textit{SAMPSEQ}=RAPID, SPARS5/10/25/50/100/200, STEP25/50/100/200/400), which describes how the non-destructive samples up the ramp are spaced and the number of samples (\textit{NSAMP}=1 to 15). The exposure time can be obtained for a particular combination from \citet[][section 13.3.6]{WFC3ProposalInfo}. \wfcsim\ is able to simulate all permitted modes, we are only limited by the available configuration files and exposure time information given.

\wfcsim 's visit planner is a function that takes all the information about the visit and calculates the exposure start times, orbit ends and buffer dumps. Specifically it takes input of \textit{NSAMP}, \textit{SUBARRAY}, \textit{SAMPSEQ}, the number of orbits and whether the telescope is scanning. It then outputs the start time of each exposure in each orbit. This involves taking into account target visibility, guide star acquisition or re-acquisition, exposure time, read time, spacecraft manoeuvres and buffer dumps from information given in \citet[][chapter 10]{WFC3InstHandbook}. The produced time sequences do not match perfectly real data or the observing schedules from the Hubble Space Telescope Astronomer's Proposal Tool\footnote{\url{http://www.stsci.edu/hst/proposing/apt}}, due to the approximate values given in the handbook. When recreating a real observation, the exposure start times from the real data can be given allowing for a more precise simulation.

\subsubsection{Non-dispersed Calibration Image}
For wavelength calibration in the analysis of \wfc\ spectroscopic data a non-dispersed (direct) image of the star is taken at the beginning of an exposure. This is used to calculate the x and y reference positions. We output a simple observation for this file consisting of a 2D Gaussian centred on the reference position to simulate the star. This allows the simulation to be analysed by existing pipelines that require this frame.

\subsection{Transit Spectroscopy: Combining Signals}

\begin{figure}
	\includegraphics[width=\columnwidth]{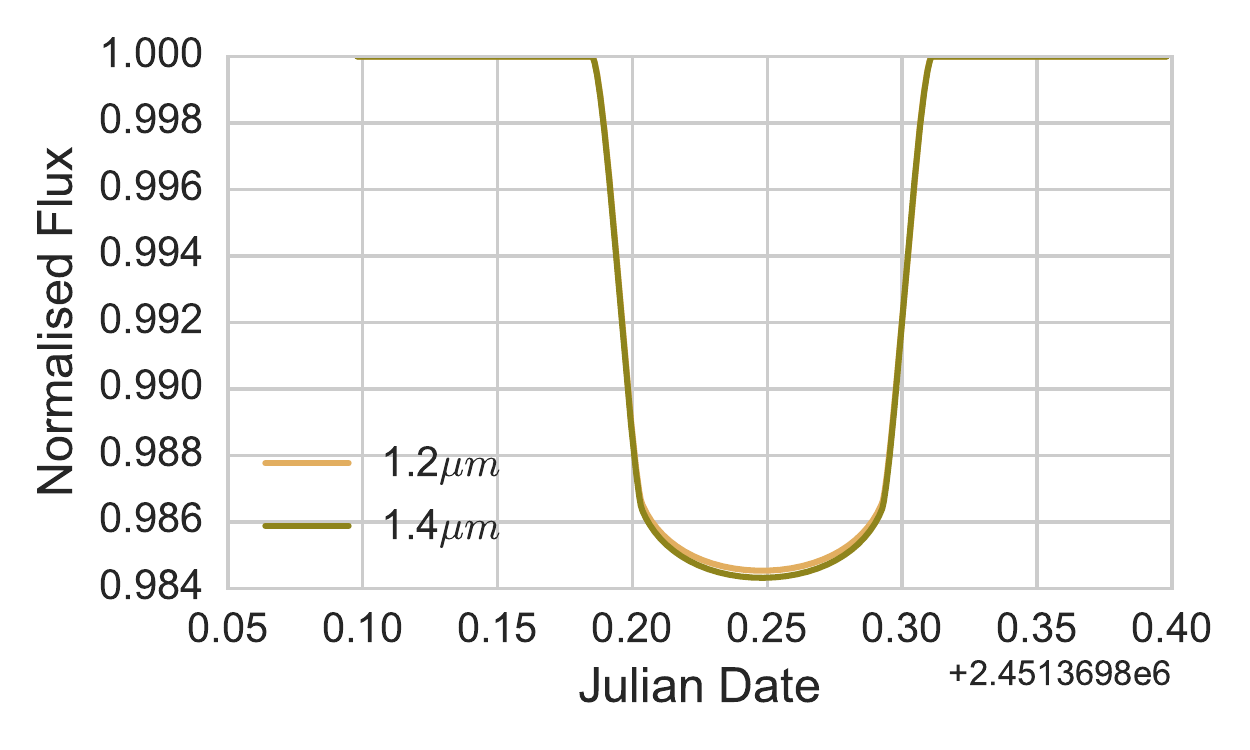}
	\caption{The light-curve model of \planet\ generated by \pylc\ using the parameters described in section \ref{sec:cs-209} at 1.2 and 1.4\,$\mu$m.}
	\label{fig:lc-model}
\end{figure}

When used in transit spectroscopy mode, the simulator must combine the planetary and stellar signals together. To do this, we need to simulate a transit event. When a planet transits a star it blocks a fraction of the stellar light we receive. As the planet moves across the stellar disc, we observe a light-curve. In order to simulate a transit we need a model of the stellar spectrum (the flux as a function of wavelengths), the planet spectrum (the transit depth as a function of wavelength) and then combine them at time $t$ and wavelength $\lambda$. A transit light curve is required for each spectral bin. They are calculated based on the above input signals and the orbital parameters of the planet, using our \pylc \footnote{\url{https://github.com/ucl-exoplanets/pylightcurve}} package (in prep.) which implements a numerical light curve model. The advantage of this implementation over other models \citep[][e.g.]{MandelAgol2002, Gimenez2006, Pal2008} is flexibility in the use of the non-linear limb darkening law \citep{Claret2000}, support for eccentric orbits, and it is written in pure Python (ensuring compatibility with the simulator). An example of the light-curve model for \planet\ (described later) is shown in Figure \ref{fig:lc-model}.

\subsection{Flux to Pixels}

The wavelength-dependent flux from the star passes through the grism and is distributed to the detector pixels. This involves scaling the flux by the grism throughput and the quantum efficiency of the detector. Then, by using the wavelength calibration coefficients of the grism we find the centre of the point-spread function (PSF) on the detector and distribute the flux to pixels based on the PSF morphology at that wavelength. We describe each of these steps in more detail below.

\subsubsection{Scaling Flux}
For the simulation we assume the input stellar flux is pre-scaled to the correct value for the \wfc\ IR channel. We provide an option to add a scaling factor which can be used to manually adjust the flux to this value. We do not, at present, attempt to implement the functionality of the Hubble Space Telescope Exposure Time Calculator.

\subsubsection{Grism sensitivity}
\begin{figure}
	\includegraphics[width=\columnwidth]{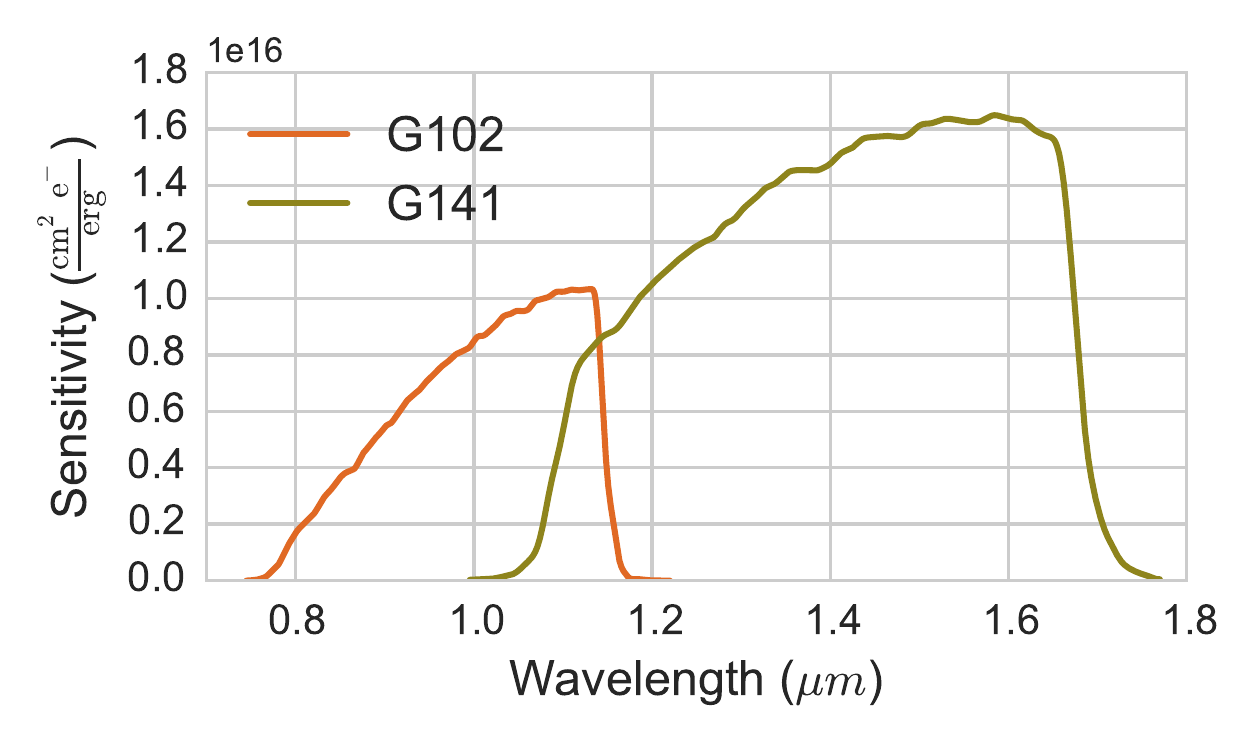}
	\caption{\wfc\ \grisms and \grisml\ grism sensitivity as calculated by \citet{WFC3IRGrismFluxCalb}. This step quantifies the conversion from flux to electrons.}
	\label{fig:calb:sensitivity}
\end{figure}

The sensitivity of the grism was calibrated by \citet{WFC3IRGrismFluxCalb} for each order of the spectrum. We only consider the first order spectrum in this version of the simulator, as it is the main scientific spectrum. The sensitivity is given in units of $\mathrm{\frac{cm^{2}\,e^{-}}{erg}}$ and quantifies the conversion of flux to electrons, accounting for both the throughput of the grism and the quantum efficiency of the detector. As shown in Figure \ref{fig:calb:sensitivity}, the sensitivity of the grism doubles between the \grisml\ wavelength limits 1.1 to 1.7\,$\mu$m. We use a linear interpolation to scale the sensitivity to the spectral bins.

\begin{figure}
	\includegraphics[width=\columnwidth]{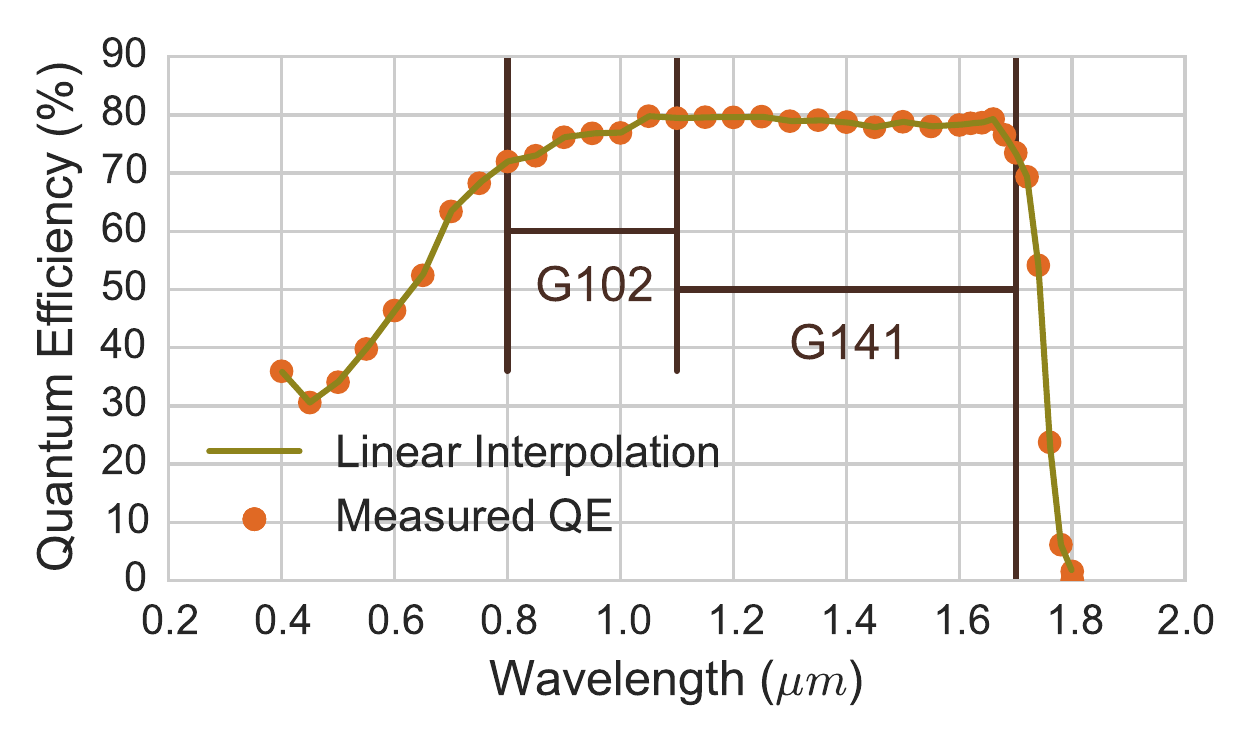}
	\caption{Quantum efficiency of the \wfc\ IR detector. Black lines indicate the edges of the wavelength coverage for the \grisms\ and \grisml\ grisms.}
	\label{fig:calb:qe}
\end{figure}

The quantum efficiency (QE) of the \wfc\ IR detector is less significant than the grism throughput but is still an important effect. The QE was calibrated on-orbit \citep{WFC3QE} and is shown in Figure \ref{fig:calb:qe}. The QE is largely flat for the \grisml\ grism but has a significant up-trend for the \grisms\ grism. We do not apply a separate QE step in our simulation as it is contained in the grism sensitivity.

\subsubsection{The field-dependent structure of the spectrum}
\label{sec:wl-sol}

When a grism is used, the source is spread out into a spectrum. The line through this spectrum is known as the trace line, while the equation giving the wavelength as a function of distance on the trace line is know as the wavelength solution. According to \citet{Kuntschner2009G102, Kuntschner2009G141} both the trace of the spectrum and the wavelength solution depend on the reference or source position on the detector (field-dependent). This is the central location of the observed star on the detector without the grism and is often obtained from a single non-dispersed (direct) image taken at the beginning of the visit. The source coordinates are indicated as $x_\text{ref}$ and $y_\text{ref}$.

\begin{figure}
	\includegraphics[width=\columnwidth]{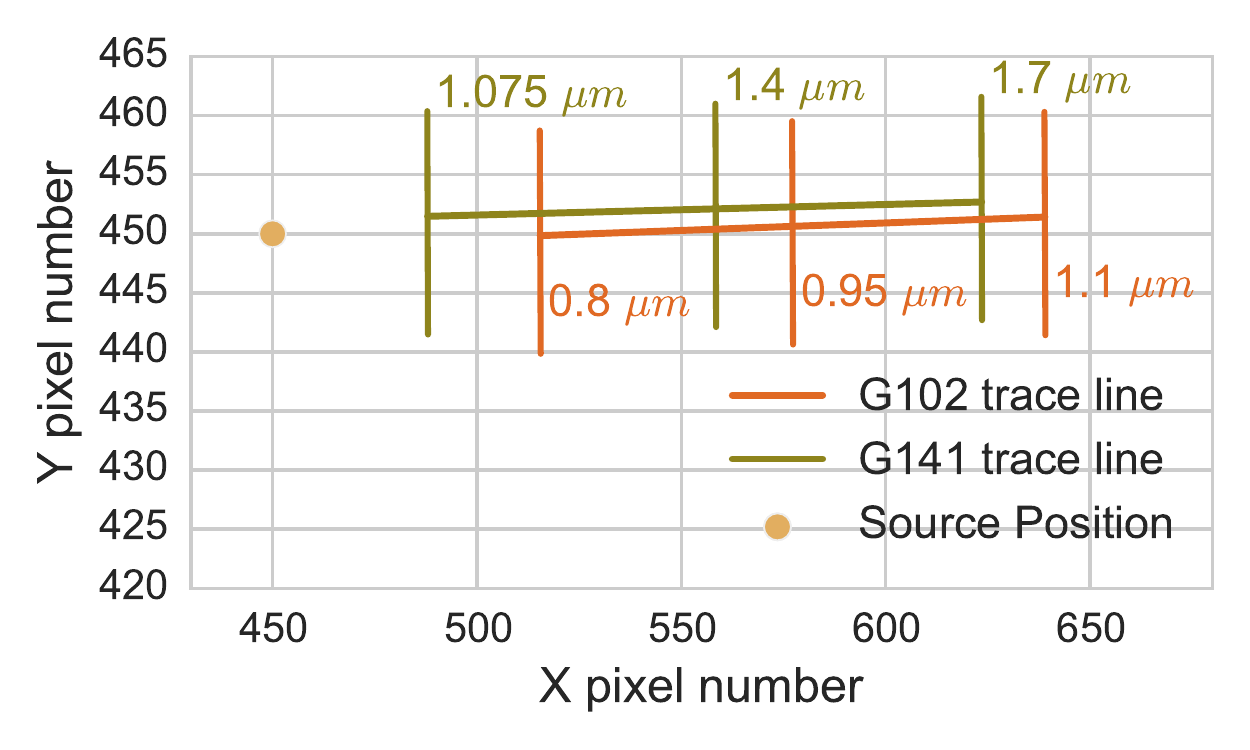}
	\caption{Trace positions of the \grisms\ and \grisml\ grisms given by the field dependent wavelength solution described in the text and a source position of (450, 450).}
	\label{fig:calb:wl-trace}
\end{figure}

Using the field-dependent wavelength solution given by \citet{Kuntschner2009G102, Kuntschner2009G141} we can map each pixel to a wavelength. An example trace for a given source position is given in Figure \ref{fig:calb:wl-trace}.

According to this calibration, both the trace of the spectrum and the wavelength solution are linear functions of $x_\mathrm{ref}$ and $y_\mathrm{ref}$. The gradient of this line is described by a 2$^\mathrm{nd}$ order polynomial while the offset described by a 1$^\mathrm{st}$ order polynomial function of $x_\mathrm{ref}$ and $y_\mathrm{ref}$. The equations describing the trace are:

\begin{eqnarray}
y = m_t x + c_t\label{eq:trace} \\
c_t = a_0 + a_1 x_{\mathrm{ref}} + a_2 y_{\mathrm{ref}} \label{eq:trace-offset}\\
\begin{split}
m_t = a_0 + a_1 x_{\mathrm{ref}} + a_2 y_{\mathrm{ref}} + a_3 x_{\mathrm{ref}}^2 + \\ a_4 x_{\mathrm{ref}} y_{\mathrm{ref}} + a_5 y_{\mathrm{ref}}^2\label{eq:trace-gradient}
\end{split}
\end{eqnarray}

Where x and y are the detector positions of the trace and the coefficients $a_n$ are defined by \citet{Kuntschner2009G102, Kuntschner2009G141} for both grisms (Table \ref{tab:grism-trace-coeff}).

The wavelength solution is given by

\begin{eqnarray}
\label{eq:hst-intro:wl-eqn}
\lambda = m_\lambda d + c_\lambda \\
c_\lambda = b_0 + b_1 x_{\mathrm{ref}} + b_2 y_{\mathrm{ref}}\\
\begin{split}
m_\lambda = b_0 + b_1 x_{\mathrm{ref}} + b_2 y_{\mathrm{ref}} + b_3 x_{\mathrm{ref}}^2 + \\ b_4 x_{\mathrm{ref}} y_{\mathrm{ref}} + b_5 y_{\mathrm{ref}}^2
\end{split}
\end{eqnarray}

where $\lambda$ is the wavelength, d is the distance between the reference and the required point along the spectrum trace and $b_n$ are the coefficients defined by \citet{Kuntschner2009G102, Kuntschner2009G141} for for both grisms (Table \ref{tab:grism-wl-sol-coeff}). 

\subsubsection{PSF}
The point-spread function (PSF) for the \wfc\ IR channel can be modelled as the linear combination of two 2D Normal distributions (N) with different standard deviations ($\sigma$): $aN_1(\sigma_1) + (1-a)N_2(\sigma_2)$. The values for the three different parameters are modelled based on the PSF ensquared energy fraction as a function of aperture size and wavelength \citep[][pg. 144]{WFC3InstHandbook}.

\begin{figure}
	\includegraphics[width=\columnwidth]{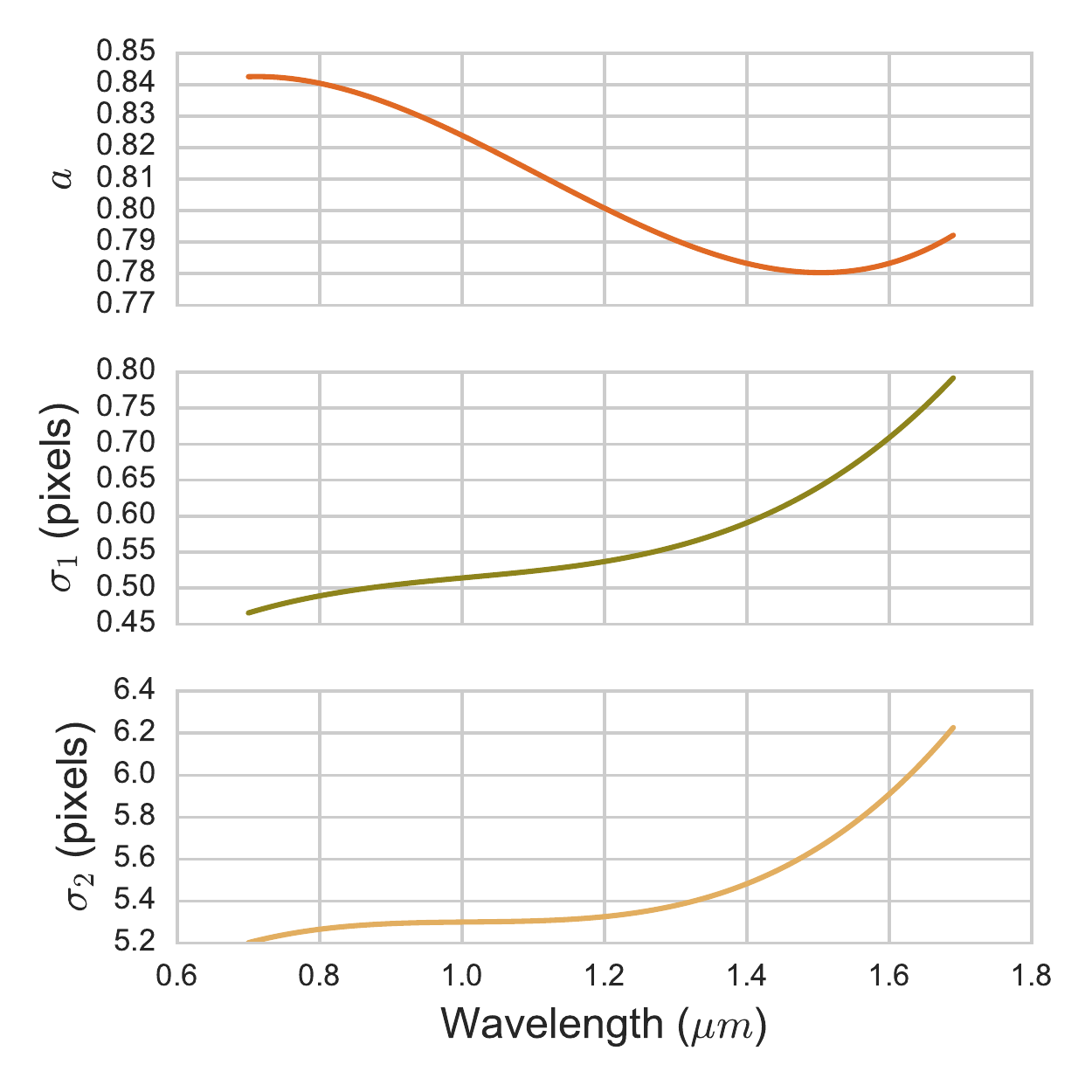}
	\caption{The PSF parameters with wavelength for the \wfc\ IR channel.}
	\label{fig:calb:psf-fwhm}
\end{figure}

The PSF parameters are shown in Figure \ref{fig:calb:psf-fwhm} as 3$^{rd}$ order polynomial functions of wavelength, a dependency that significantly affects the final spectrum (for the coefficients used see Table \ref{tab:psf-coeff}). A comparison between the reported \hst\ values and our simulation at 0.9 and 1.4\,$\mu$m can be found in Table \ref{tab:calb:psf-comparison}. The difference is due to the diagonal components of the \hst\ PSF, which we do not include. However, the size of the PSF wings is comparable with the real data.

\begin{table}
    \small
    \center
    \caption{\hst\ and \wfcsim\ PSF ensquared energy fraction as a function of aperture size (pixels) at 0.9 and 1.4\,$\mu$m.}
    \label{tab:calb:psf-comparison}
    \begin{tabular}{c | c c | c c }
        \hline \hline
        aperture size			& \multicolumn{2}{c |}{0.9\,$\mu$m}		& \multicolumn{2}{c}{1.4\,$\mu$m}		\\
        (pixels)				& \hst\	&	\wfcsim				& \hst\	&	\wfcsim				\\ [0.1ex]
        \hline
        1$\times$1			& 0.398	& 0.391					& 0.314	& 0.278					\\
        3$\times$3			& 0.802	& 0.839					& 0.706	& 0.775					\\
        5$\times$5			& 0.858	& 0.857					& 0.830	& 0.811					\\
        7$\times$7			& 0.894	& 0.875					& 0.855	& 0.832					\\
        9$\times$9			& 0.917	& 0.896					& 0.888	& 0.859					\\
        11$\times$11			& 0.934	& 0.916					& 0.905	& 0.886					\\
        13$\times$13			& 0.946	& 0.935					& 0.917	& 0.909					\\
        15$\times$15			& 0.955	& 0.952					& 0.931	& 0.931					\\
        17$\times$17			& 0.962	& 0.966					& 0.942	& 0.949					\\
        19$\times$19			& 0.966	& 0.976					& 0.949	& 0.965					\\
        21$\times$21			& 0.969	& 0.985					& 0.956	& 0.975					\\
        23$\times$23			& 0.971	& 0.991					& 0.961	& 0.984					\\
        25$\times$25			& 0.973	& 0.995					& 0.965	& 0.990					\\
    \end{tabular}
\end{table}

The final step is to distribute on the detector the total number of electrons for each wavelength bin, given the centre and the shape of the PSF at this wavelength. We associate to every electron two random numbers drawn from one of the two normal distributions that construct the PSF -- a fraction of 1/a of all the electrons belong to $N_1$ and a fraction of 1/(1-a) belong to $N_2$. These random numbers are then added to the centre of the PSF to give the x and y pixel coordinates of all the electrons. To optimize this process, the random number generation is implemented in C, using the Box-Muler method, and is accessed by Python using the Cython compiler (\url{http://cython.org/}). 

\begin{figure}
\includegraphics[width=\columnwidth]{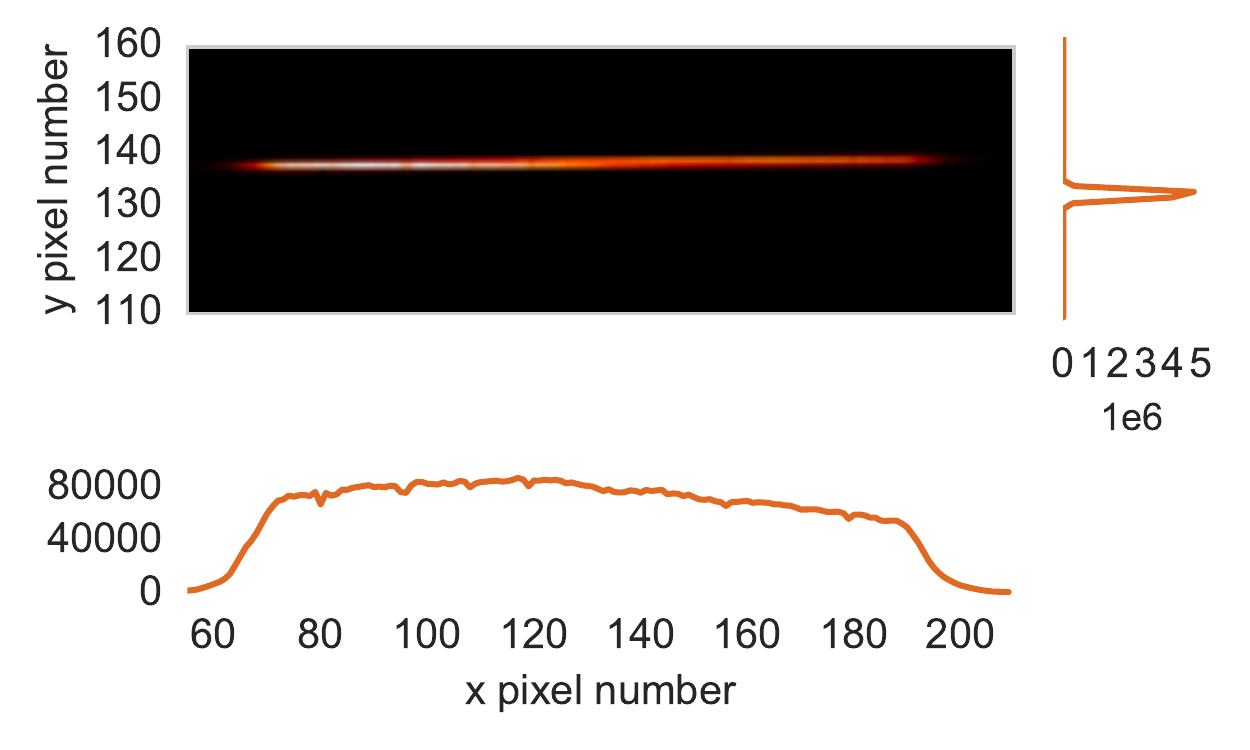}
\caption{Cropped exposure of a simulation of a staring mode frame for \planet\ using \wfcsim\ with the summed counts of the frame in both axes shown. The figure plots the difference between the last and the first reads of the exposure.}
\label{fig:staring-mode-frame}
\end{figure}

At this point, we have described all the parts needed to form a simple staring mode frame. Using the parameters described in section \ref{sec:cs-209} we have generated a staring mode frame shown in Figure \ref{fig:staring-mode-frame}.

\subsection{Noise and data reductions}

There are many different reductions needed for processing \wfc\ data which are fairly well calibrated from on-orbit calibration studies. Noise sources have been quantified in a similar way. We describe here how we implement each of these steps in the simulation. Note that in our figures we show calibrations for the 256 subarray as this is one of the most commonly used for exoplanet transmission spectroscopy with \wfc.

\subsubsection{Photon Noise}
Photon noise is added by converting the stellar flux rate for each sub-sample into counts using a Poisson distribution before distributing the flux into pixels.

\subsubsection{Flat field}
\begin{figure}
	\includegraphics[width=\columnwidth]{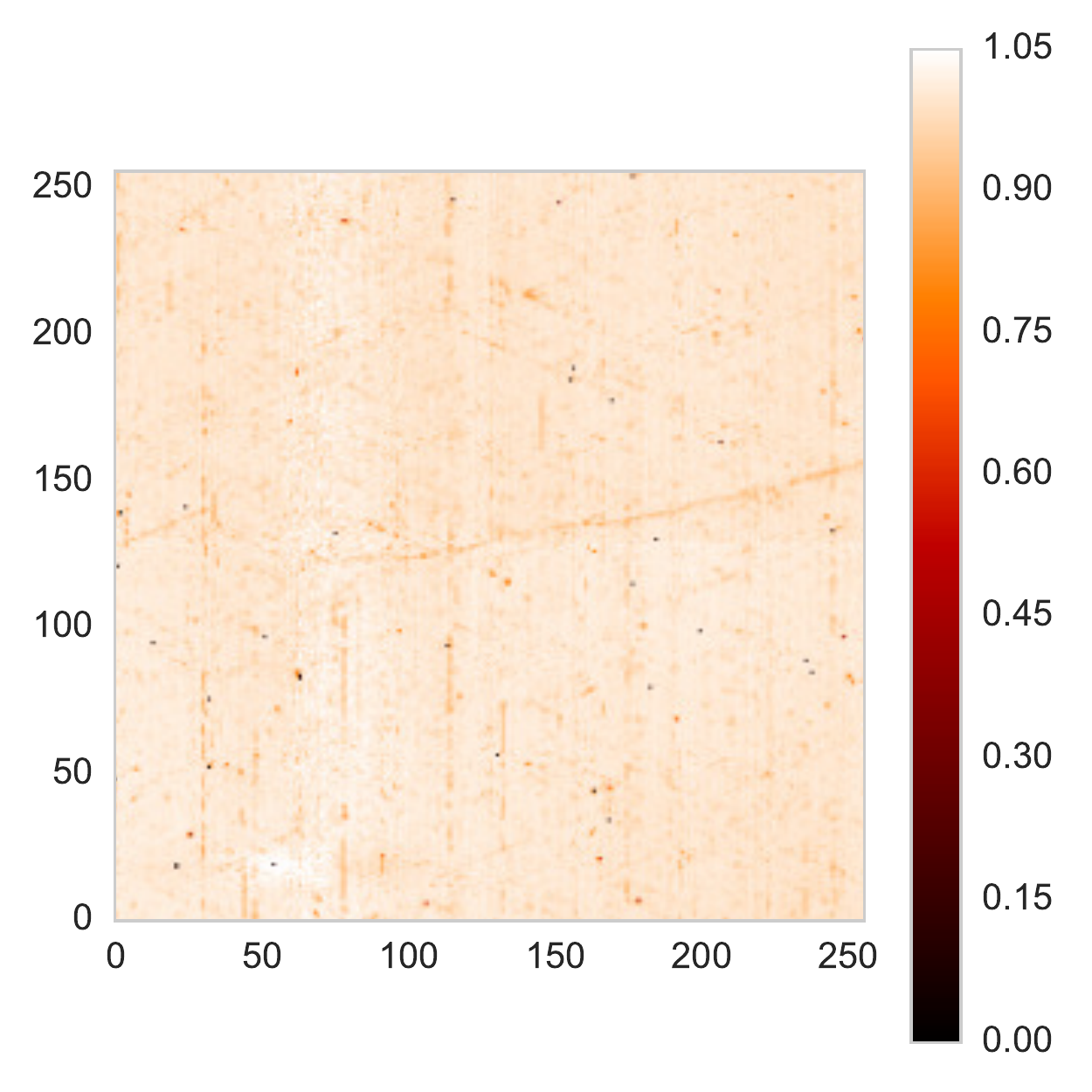}
	\caption{Flat Field cube from \citet{WFC3FluxCalib} for the \grisml\ grism and 256 subarray.}
	\label{fig:calb:flat}
\end{figure}

For the flat field we use the flat-field cube described in  \citet{WFC3FluxCalib} (Figure \ref{fig:calb:flat}). The flat field for grism spectroscopy is a 3$^\mathrm{rd}$ order polynomial with the coefficients given in the flat field cube. The flat is calculated on a per subsample basis by determining the wavelength of every pixel at the subsample reference position.

\subsubsection{Sky Background}
\begin{figure}
	\includegraphics[width=\columnwidth]{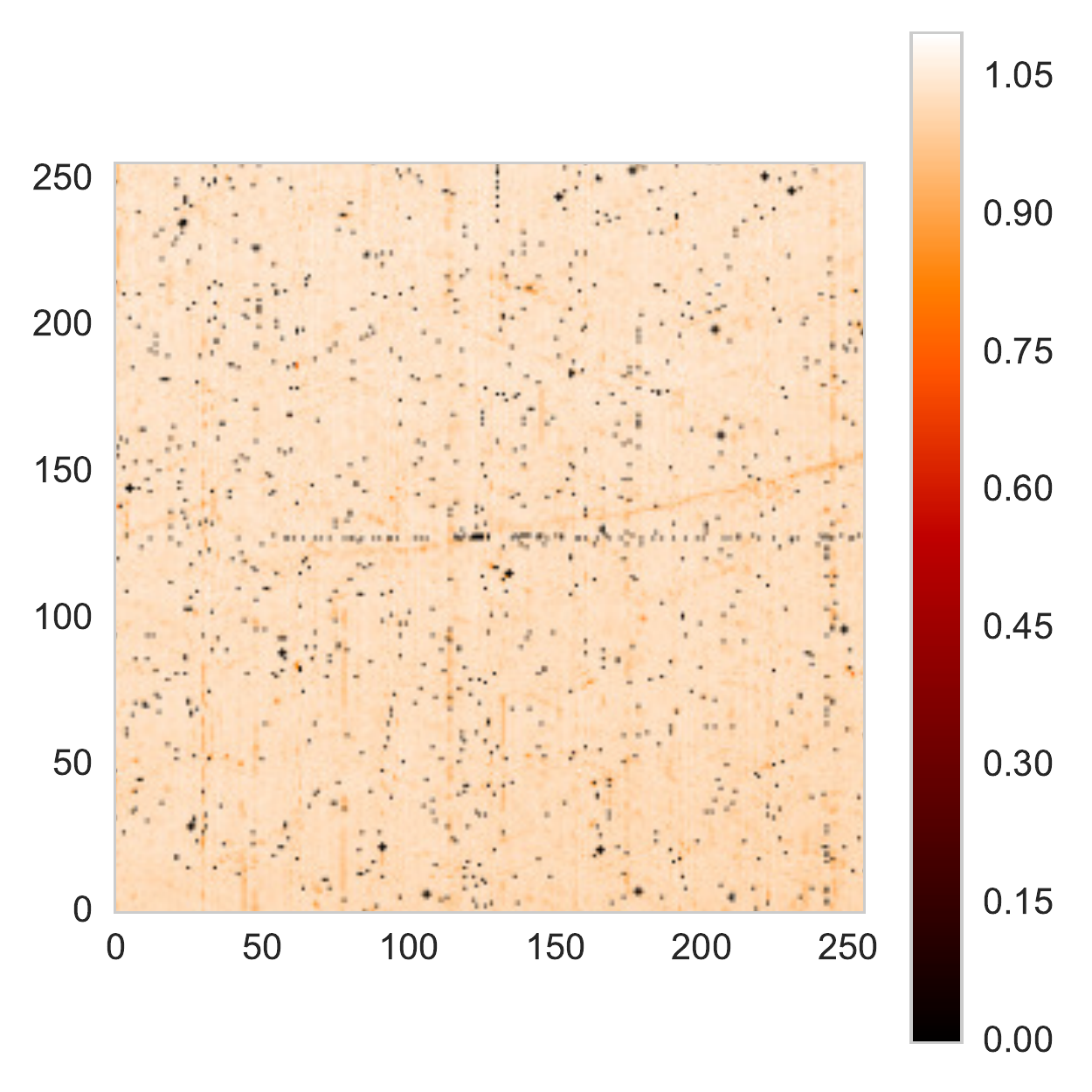}
	\caption{Master sky image from \citet{WFC3MasterSky} for the \grisml\ grism and 256 subarray.}
	\label{fig:calb:sky}
\end{figure}

The sky background is generated by multiplying the master sky image \citep{WFC3MasterSky} (Figure \ref{fig:calb:sky}), by a factor depending on the target and then sampling each pixel from a Poisson distribution to add noise. We acknowledge the recent source-dependent master sky images by \citet{WFC3MasterSky2015} and plan to implement them in a future version.

\subsubsection{Cosmic Rays (CR)}
Cosmic rays impact the detector at a rate of 11\,$\pm$\,1.5\,s$^{-1}$ across the full frame \citep[][pg 153]{WFC3InstHandbook}. We add cosmic rays to each sample by scaling the rate by the exposure time and subbarray size and sample a Poisson distribution with this rate to give the number of impacts. We then randomly choose a pixel for the impact to occur and the number of counts between 10--35\,kDN to add to that pixel.

\subsubsection{Gain variations}
The gain is the $e^-/DN$ conversion of the amplifier used to read a pixel. This is composed of a constant factor 2.35 and gain variations per quadrant which are specified in the gain calibration file \textit{u4m1335mi\_pfl.fits}. We therefore divide the simulation counts by 2.35 and then multiply by the gain variations.

\subsubsection{Dark Current}
\begin{figure}
	\includegraphics[width=\columnwidth]{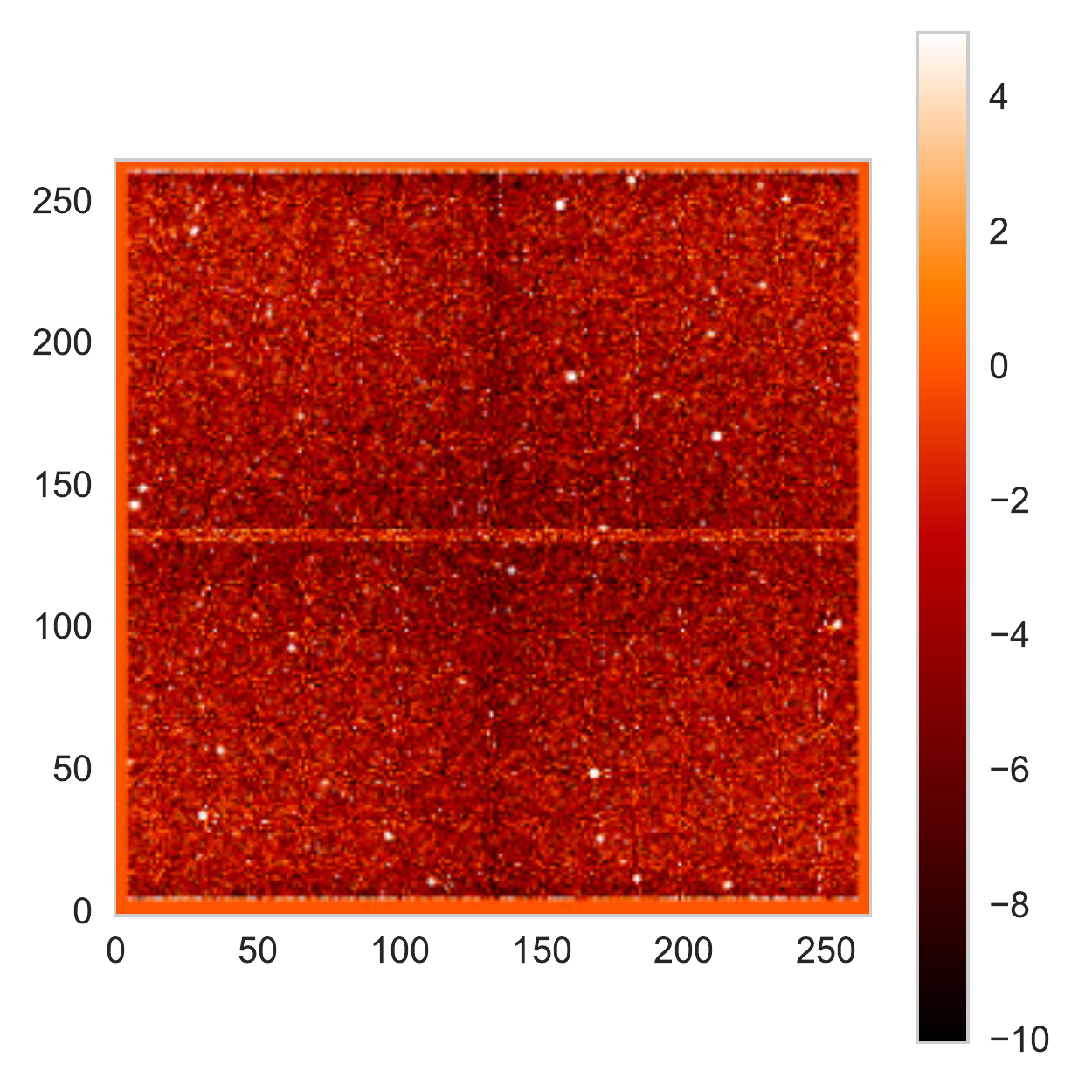}
	\caption{\wfc\ IR superdark \textit{u4819493i\_drk.fits} for a 256 subarray at NSAMP 5 in SPARS10 mode covering 29.6\,s. The frame is scaled between -10 and 5 electrons per pixel. The mean is -1.65, the median is -3.37, the minimum is -14.3, and the maximum is 14000 electrons per pixel. This includes any bad pixels (cold and hot).}
	\label{fig:calb:dark}
\end{figure}

Dark current rates are typically around 0.05 e$^-$s$^{-1}$ per pixel, which is at negligible level for most purposes \citep{WFC3SuperDark2014}. Despite this, for completeness we add dark current, using the superdarks \citep{WFC3SuperDark2014}, created by averaging many dark exposures from cycles 17, 18, 19 and 20. They are specific to the subarray and readout mode used, as the dark current scales non-linearly with time. An example superdark is shown in Figure \ref{fig:calb:dark}. Dark current is added to each sample at the end of the simulation on a per read basis.

\subsubsection{Non-linearity}
The \wfc\ IR detector reaches 5\% non-linearity at 70,000\,e$^-$ or $\sim$ 28,000\,DN with a maximum well depth of 40,000\,DN \citep{WFC3OldNonLinear}.

The form of the non-linearity is described by \citet{WFC3OldNonLinear} on a per quadrant basis. We note the more precise non-linearity correction per pixel by \citet{WFC3NewNonLinear} but the coefficients to implement this correction were not readily available. The per quadrant version is described by a 3$^\mathrm{rd}$ order polynomial with the coefficients \textit{c} given in the calibration file. The conversion from the non-linear $F$ to the linear case $F_\mathrm{c}$ is then:

\begin{equation}
F_\mathrm{c} = (1+c_1 + c_2 F + c_3 F^2 + c_4 F^3) F
\end{equation}

For our simulations, we want the reverse, i.e. to convert linear counts to non-linear. To do this, we use the Newton-Raphson numerical method to find the root of the above equation that is closest to the original counts.

\subsubsection{Count level clipping}
As the non-linearity correction only applies to counts up to the 5\% non-linearity limit we clip the maximum values of all pixels to this level (70,000\,e$^-$) and consider them saturated.

\subsubsection{Bias and bias drifts}
\label{sec:bias}
\begin{figure}
	\includegraphics[width=\columnwidth]{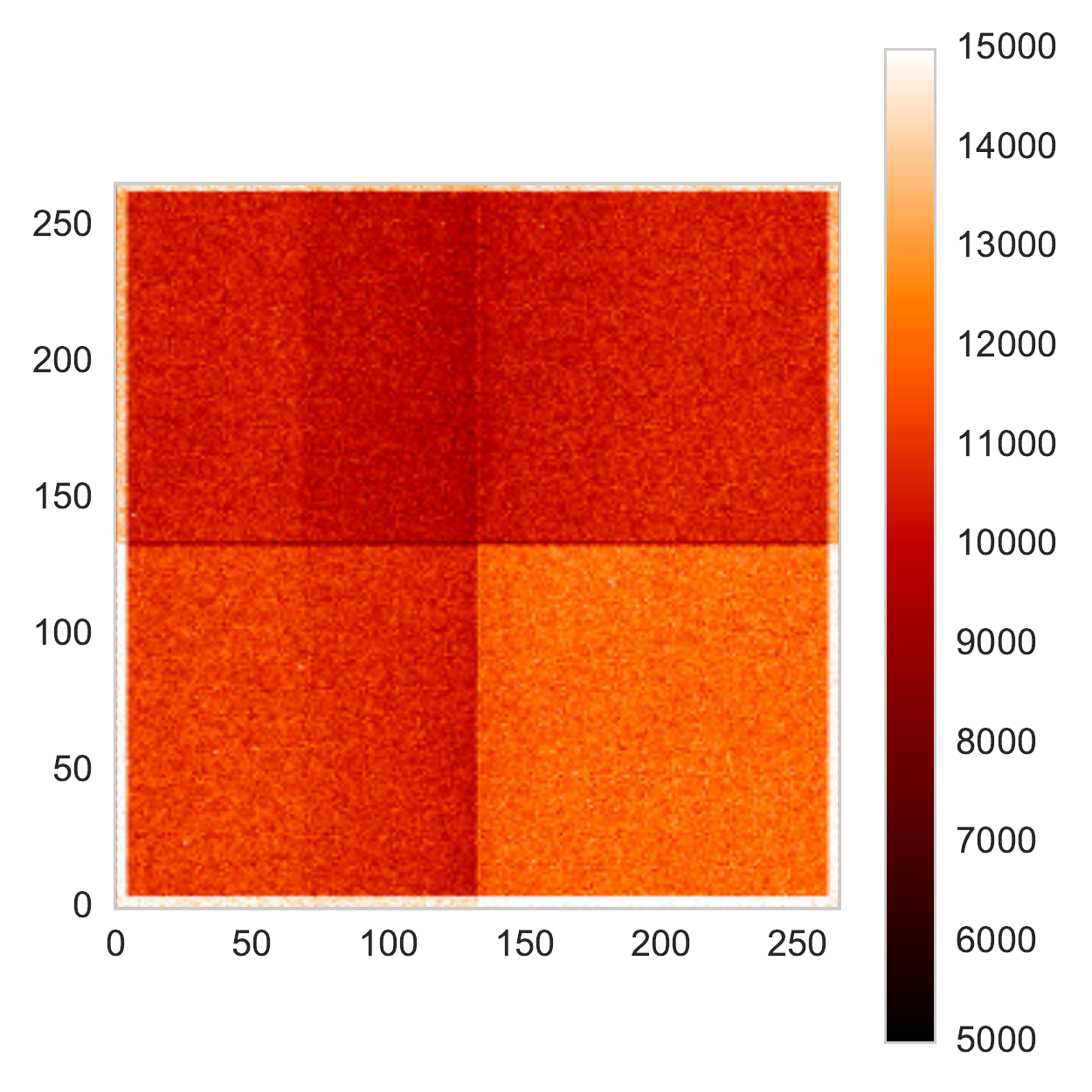}
	\caption{The initial bias frame for the 256 subarray mode used for the zero-read. The frame is included mostly for consistency with real data as it is removed in zero-read subtraction - the first stage of analysis. The scale has been clipped between 5,000 and 19,000\,DN.}
	\label{fig:calb:initalbias}
\end{figure}

When the detector is reset there is an initial bias level of around 11,000\,DN with small differences per quadrant. This bias level gives the look of a raw \wfc\ frame before subtracting the zero-read and is nearly fully corrected by subtracting the zero-read \citep[][pg. 126]{WFC3DataHandbook}. For consistency with real data we include this effect by creating an initial bias file from the zero-read of two frames of real data obtained from the \stsci\ \mast \footnote{\url{https://archive.stsci.edu}}. Specifically we use the upper half of the file \textit{icdp02b0q\_raw.fits} and lower half of the \textit{icdp02b1q\_raw.fits} file (to obtain areas not contaminated by the spectrum). The resulting file is shown in Figure \ref{fig:calb:initalbias}.

The bias level drifts during an exposure. The \wfc\ IR detector allows correcting for this effect by having a five pixel outer border around the 1014\,$\times$\,1014 array that is not sensitive to incoming light. These are known as reference pixels and are a proxy to the drifting bias level. By averaging them and subtracting the result from each read we can correct the bias drift. We do not attempt to implement bias drifts in the current simulation, instead we set all the reference pixels to 0 before applying the read noise.

\subsubsection{Read Noise}
Read noise for the \wfc\ IR channel is Gaussian with a standard deviation between 19.8 to 21.9\,e$^-$ \citep{WFC3ReadNoise} after subtracting the zero-read. For a single read, we therefore sample a normal distribution for each pixel with standard deviation of 20/$\sqrt2$ = 14.1\,e$^-$ (6\,DN) and apply it non-cumulatively to each sample. 

\subsection{Generating an Exposure}

\subsubsection{Spatial Scanning}
Spatial scanning mode is created by combining multiple staring modes frames together. We generate subsamples at the subsample rate, typically 5--30\,ms depending on the scanning speed. The planetary and stellar spectra are combined per subsample so the transit is progressing throughout the scan. Trends such as the scan speed variations are applied on a subsample basis.

\subsubsection{Reading the array} \label{reading}
The \wfc\ IR detector is read separately in quadrants with each quadrant read towards the centre of the frame. This process takes 0.278\,s in total for the 256 subarray meaning that the flux content of the different pixels is not recorded at the same time. For all pixel exposure times to be equal, the zero-read must be subtracted. 

The read time and direction has an impact when scanning due to the up-stream / down-stream effect identified by \citet{WFC3ScancCalib2012}. This causes a change in exposure time depending on whether the scan is with or against the read direction. We do not simulate this effect at present but plan to include it in a future version as it has a particular effect when a visit is composed of scans in both directions \citep[see][]{Knutson2014a, Tsiaras2016}).

\subsubsection{Samples up the ramp}
During an exposure, the detector is  read several times, in what are known as `samples up-the-ramp' which we refer to as reads. We generate each read separately to recreate this data structure. The zero-read is currently implemented as the initial bias described in Section \ref{sec:bias} with read noise.

\subsubsection{Output to FITS format}

The simulations output to FITS files in the same format as \hst\ \wfc\ data and contain most of the same header information. In addition we include several new header keys giving details about the simulation such as the simulator version, x-shift amount and the planet and star parameters used. A full list is given in table \ref{tab:fits-headers}.

\subsection{Systematics}

Data obtained with \wfc\ include several systematics that can be measured per exposure over an entire visit. These include the shift in the starting x and y position in each frame (x-shifts and y-shifts), scan speed variations and the `hook' or `ramp' effect.

\subsubsection{Reference position shifts (x,y)}
Shifts in the reference position ($x_\text{ref}$, $y_\text{ref}$) of the spectrum on the detector are seen in real data in both the x and y axis as small shifts in the spectrum position. x-shifts typically have a total amplitude of 0.1 to 1 pixel per visit \citep{Deming2013, Tsiaras2015} and are approximately linear with time. We implement x-shifts using a function which generates the shift in $x_\text{ref}$ position for each exposure, by default this is a linear trend. For matching real data the position of $x_\text{ref}$ can be given on a per frame basis.

The y-shifts are implemented in the same way as x-shifts but are often considered negligible in real data as the total shift is commonly only $\sim 0.1$ pixels \citep{Tsiaras2015}. However, there are examples where the shift is higher, i.e. a total of $\sim 1.3$ pixels in \citet{Tsiaras2016} where y-shifts can not be ignored.

In addition to the position shifts from one frame to an other jitter noise is also present in \hst\ observations. Jitter occurs due to the imperfect pointing of the telescope during an observation, and results in an effective increase in the size of the PSF. Jitter noise, across both axis, is included as uncertainty in the reference position ($x_\text{ref}$, $y_\text{ref}$) of the spectrum. To match real data sets, the input value of the jitter noise can be derived from the available jit files. For \hst, the jitter is typically between 0.003 and 0.005$''$ (0.022 to 0.037 pixels). For the \planet\ data set used in section \ref{sec:cs-209} the deviation from a straight line scanning trajectory, which is assumed in \wfcsim, is $\pm 0.0013''$ (0.01 pixels). However, these measurements represent time intervals of 3 seconds, corresponding to $0.0013 \sqrt{3} = 0.0023''$ (0.017 pixels) for intervals of 1 second. In section \ref{sec:cs-209} we used a jitter of 0.02 pixels.

\begin{figure}
\centering
\includegraphics[width=\columnwidth]{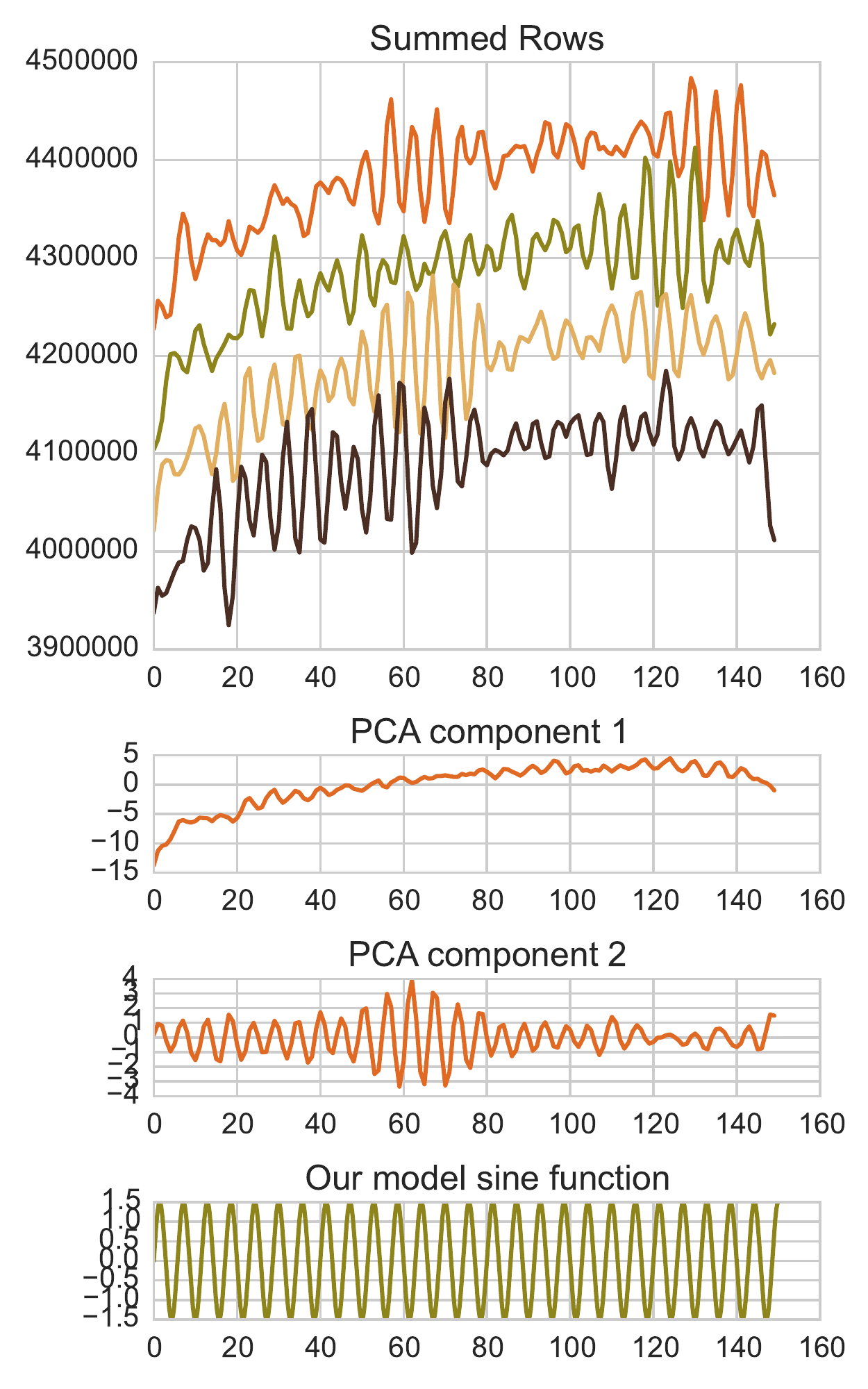}
\caption{Measuring scan speed variations from orbit 3 of \planet. The upper plot shows the first 4 signals, created by summing the rows of each exposure (offset for clarity). The mid plots show the first 2 principle components from a PCA of all 20 signals. The lower plot shows our sine model of the general SSV trend.}
\label{fig:calb:ssv}
\end{figure}

\begin{table*}
	\small
	\center
	\caption{Summary of simulated effects.}
	\label{tab:effects}
	\begin{tabular}{c | c }
		\hline \hline
		Flux distribution / positioning			& Detector effects							\\ [0.1ex]
		\hline
		\multicolumn{2}{c}{included}													\\
		\hline	
		flux-to-electron conversion			& zero-read								\\
		random sampling of the PSF			& read noise								\\
		wavelength-dependent PSF			& dark current								\\
		field-dependent spectrum trace			& gain conversion							\\
		field-dependent wavelength solution		& non-linearity								\\
		spatial scanning					& wavelength-dependent flat-field				\\
		scan speed variations				& field-dependent sky background				\\
		position shifts						& cosmic rays								\\
		\textbf{jitter noise}					& 										\\
		\hline
		\multicolumn{2}{c}{not included}												\\
		\hline
		``blobs''							& image persistance 							\\
		satellite trails						& cross talk								\\
										& inter-pixel capacitance						\\
										& charge diffusion							\\
										& non-ideal A/D converters (Section \ref{sec:bias})	\\
										& ``snowballs''								\\
										& up-stream/down-stream effect (Section \ref{reading})	\\

	\end{tabular}
\end{table*}

\subsubsection{Scan speed variations (SSVs)}

If the FGS is used for controlling the scan there is a flicker which appears as a wave-like variation in flux in the direction of the scan, due to FGS feedback \citep{WFC3ScancCalib2012}. 

We looked at SSVs in the \planet\ data set used in section~\ref{sec:cs-209}. After applying all reductions to the frames we choose a single orbit (orbit 3) and summed the rows of the spectrum to give the total flux per row in the scan direction for each of the 20 frames. Visually these variations appear to be modulated sine waves (see Figure \ref{fig:calb:ssv} top). We confirmed this behaviour by performing Principle Component Analysis on these 20 summed rows and recovered the first 2 principal components which appear to be 1) the general increase in flux across the rows 2) a sine wave with varying amplitude (see Figure \ref{fig:calb:ssv} middle). We fit the second principle component to give an approximate sine model of period=0.7\,seconds, standard deviation=1.5 and phase=0 (see Figure \ref{fig:calb:ssv} bottom). These values are configurable in the simulation to adapt to other data sets.

The sinusoidal model is applied to the simulation by changing the exposure time of each subsample to that defined by the sine model with mean equal to the subsample time. We note that the real scan speed variations are reasonably complex, appear to modulate and have occasional large `blips' in the data consisting of a large peak and trough in the flux level. We only give an initial approximation of the effect for this data set here as an in depth analysis is beyond the scope of this paper.

\subsubsection{Hook and long term ramp}

There are two more trends often seen in real data which are often combined into a single reduction step. This is the hook effect which appears to reset after each buffer dump and a long term ramp which is a gradual decrease in the level of flux with time across a visit.

The `hook' was first described by \citet{Berta2012} and implemented as a exponential fit by \citet{Kreidberg-gj1214b}. We adopt the form of \citet{Tsiaras2015} (Equation \ref{rampfunction}) and apply it as a scaling factor to the flux for each subsample.

\begin{equation}
	R(t) = (1 - r_a (t-t_v))(1-r_{b1} e^{-r_{b2} (t-t_o)})
	\label{rampfunction}
\end{equation}

Where $t$ is time, $t_v$ the time when the visit starts, $t_o$ the time when the orbit in which the frame belongs starts, $r_a$ the slope of the linear long-term `ramp' and ($r_{b1},r_{b2}$) the coefficients of the exponential short-term `ramp'. The values of these coefficients can be set in the configuration to match a similar observation.

\subsection{Summary of simulated effects}

Table \ref{tab:effects} summarizes the effects related to the spatial scanning technique and the WFC3 IR detector, and here we discuss the effects that are not currently  implemented in \wfcsim.

\paragraph{``Bolobs''}

The ``blobs'' are areas of 10 to 15 pixels that absorb up to 15\% of the incoming light at their centers. The behaviour of the ``blobs'' appears to be date-dependent, making their implementation in \wfcsim\ more difficult than the offer effects \citep[][pg. 160]{WFC3InstHandbook}.

\paragraph{Satellite trails}
Such trails can appear when satellites cross the observed field of view. Due to their brightness, satellites saturate a number of pixels (approximately 10 to 15), depending on the length of the trail. Saturation can also affect consecutive observations because of image persistence. In the case of exoplanetary observations, the exposure times are short and the satellite trails are more rare.

\paragraph{Image persistance}

Due to image persistence, an exposure of the detector is causing ghost images or afterglows in the consecutive ones \citep[][pg. 154]{WFC3InstHandbook}. This effect is, possibly, the cause of the ``hook'' seen in each \hst\ orbit. Implementing this effect requires the coupling between different exposures, which is, currently, not available in \wfcsim.

\paragraph{Crosstalk}
Is the effect when a bright source in one quadrant is causing electronic ghosting in another quadrant, coupled to the first one. In the WFC3/IR channel, the two quadrants on the left side of the detector are coupled, and the the two quadrants on the right side are also coupled. However, this effect is below the background noise for unsaturated sources \citep[][pg. 66]{WFC3InstHandbook}.

\paragraph{Inter-pixel capacitance (IPC) and charge diffusion (CD)}
Both effects cause a fraction of the charge collected by an individual pixel to ``leak'' into its neighboring pixels and therefore the recorded image to appear smoother. We plan to include both effects in the future, by using appropriate smoothing kernels. IPC is a form of crosstalk and CD is caused during charge transfer through the detector, hence, only the later is wavelength-dependent \citep{WFC3IPC2011}.

\paragraph{``Snowballs''}
``Snowballs'' have the same behavior as the cosmic rays, but they affect about 10 pixels and they contain between 200,000 and 500,000 e$^-$. Because they appear only about once per hour over the entire WFC3 detector, this effect is not common in spatially scanned data sets \citep{WFC3Snowballs2015}.

\section{\planet: a case study}
\label{sec:cs-209}

\begin{table}
    \small
    \center
    \caption{Input values for the \system\ system parameters and ramp coefficients (used in Equation \ref{rampfunction}) compared to fitting results on the white light curve ($1.125-1.65 \, \mu \mathrm{m}$).}
    \label{tab:parameters}
    \begin{tabular}{c | c | c }
        \hline \hline
        parameter					& input value	& retrieved values				\\ [0.1ex]
        \hline
        $T_0$ [HJD--2456196]		& 0.28835		& 2456196.28834$\pm$0.00002	\\
        $R_\mathrm{p}/Rs$ [mean]	& --			& 0.12077$\pm$0.00003			\\
        Period [days]				& 3.52474859	& --							\\
        $a/R_*$					& 8.8587		& --							\\
        $i$ [deg]					& 86.71		& --							\\
        $r_a$					& 0.005		& 0.005$\pm$0.0001				\\
        $r_{b1}$					& 0.0011		& 0.0011$\pm$0.00002			\\
        $r_{b2}$					& 400		& 395$\pm$12					\\
    \end{tabular}
\end{table}

\begin{figure}
	\centering
	\includegraphics[width=\columnwidth]{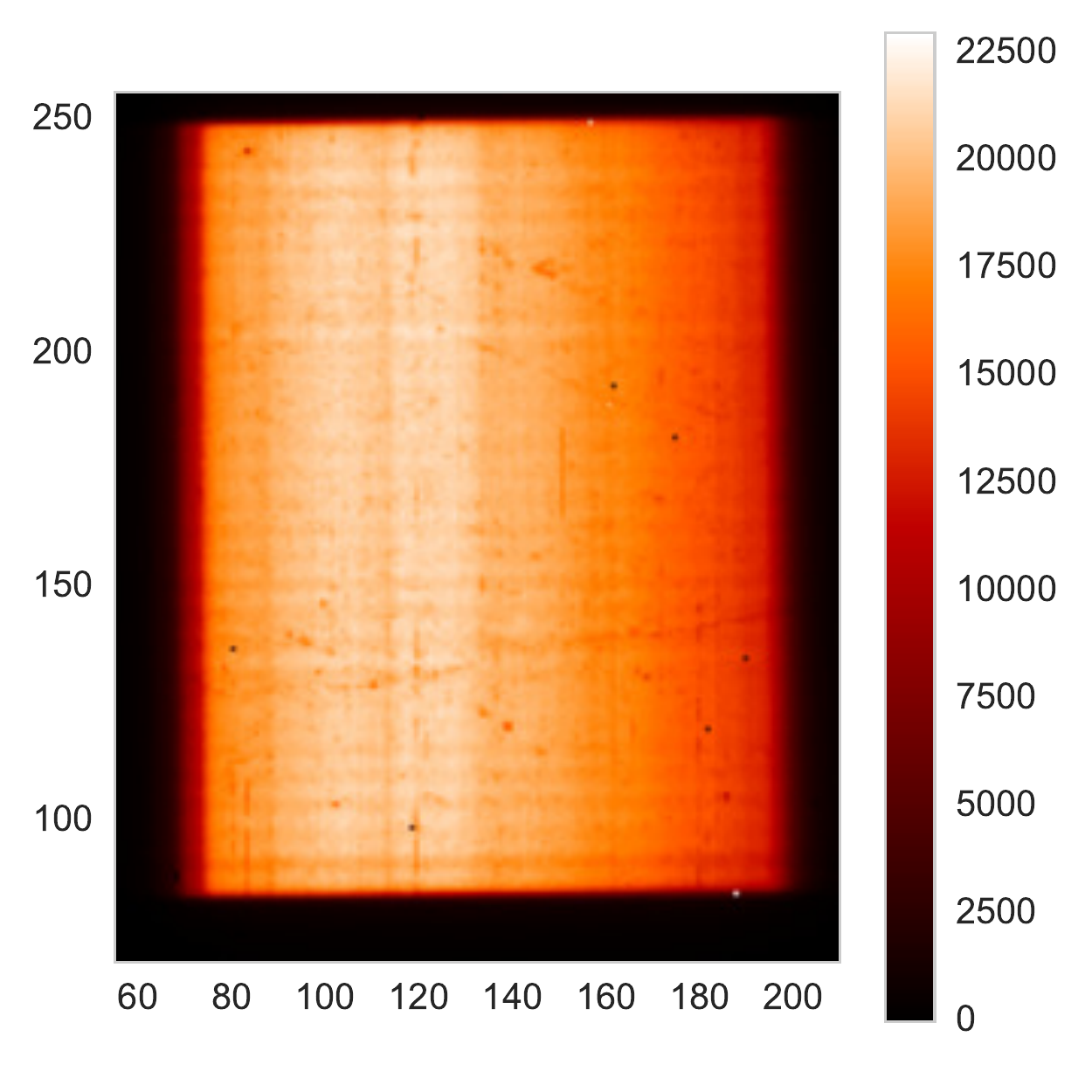}
	\includegraphics[width=\columnwidth]{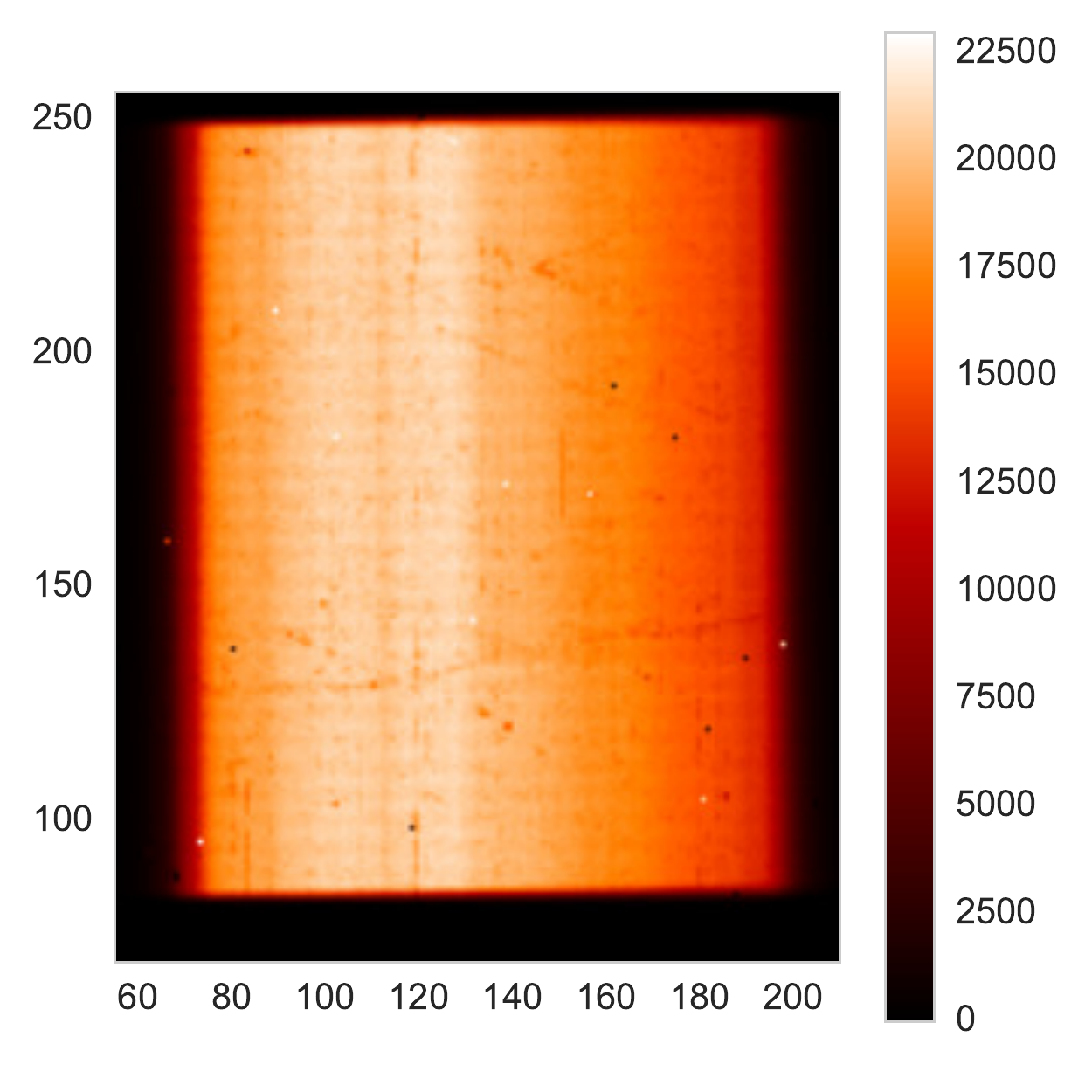}
	\label{fig:209:raw_frame_comparison}
	\caption{Comparison of a real raw frame (ibh726meq\_raw.fits) of \system\ from proposal 12181 (upper) and the \wfcsim\ simulation (lower). Both represent the flux difference between the last and the first reads and are cropped to the same limits to highlight the spectrum. The main visual difference is the non-uniform scan speed variations in the real frame. Simulated cosmic rays are random and will differ from the real frame.}
\end{figure}

As an application, we simulate the scanning-mode spectroscopic observations of the hot Jupiter \planet\ (ID: 12181, PI: Drake Deming). The simulated data set consists of 83 images and each image of 5 up-the-ramp samples with a size of 266 $\times$ 266 pixels in the SPARS10 mode. As a result, the total exposure time, maximum signal level and total scanning length are very similar to the real frames (22.32 seconds, $4.8 \times 10^4$ electrons per pixel and ~170 pixels, respectively). Our simulation also includes one non-dispersed (direct) image of the target, at the beginning of the observation, for calibration reasons, similar to the real data set. A raw frame from the simulator and real data is shown in Figure \ref{fig:209:raw_frame_comparison}.

For this simulation we use a PHOENIX model \citep{PHOENIXBTsett, Baraffe2015} for the stellar spectrum for a star as similar as possible to \system\ ($T_\mathrm{eff} = 6065 \, \mathrm{K},  \mathrm{[Fe/H]} = 0.00 \, \mathrm{[dex]},  \mathrm{log}(g_*) = 4.361 \, \mathrm{[cgs]}$ and a planetary spectrum generated by $\mathcal{T}$-REx \citep{TaurexTransmission} for \planet\ (including $\mathrm{H}_2\mathrm{O}$, $\mathrm{NH}_3$, HCN and clouds). The sky background, x and y shifts and exposure times are configured to match those recovered by \citet{Tsiaras2015}.

The total execution time for a single spatially scanned image is of the order of 37 seconds, but it is highly depended on the hardware. Among the different stages of the simulation, 16\% of the time is spent on calculating the flux as a function of wavelength, 68\% on randomly distributing the flux into the pixels, 9\% on applying the wavelength-dependent flat-field and 7\% on all the other processes.

\newpage

\subsection{Reduction-Calibration-Extraction} \label{subsub:rce}

We analyse the simulated data set using our specialised pipeline for reduction, calibration and extraction of scanning-mode spectroscopic data \citep{Tsiaras2015} for a one-to-one comparison with the real data set.

\begin{figure}
\centering
\includegraphics[width=\columnwidth]{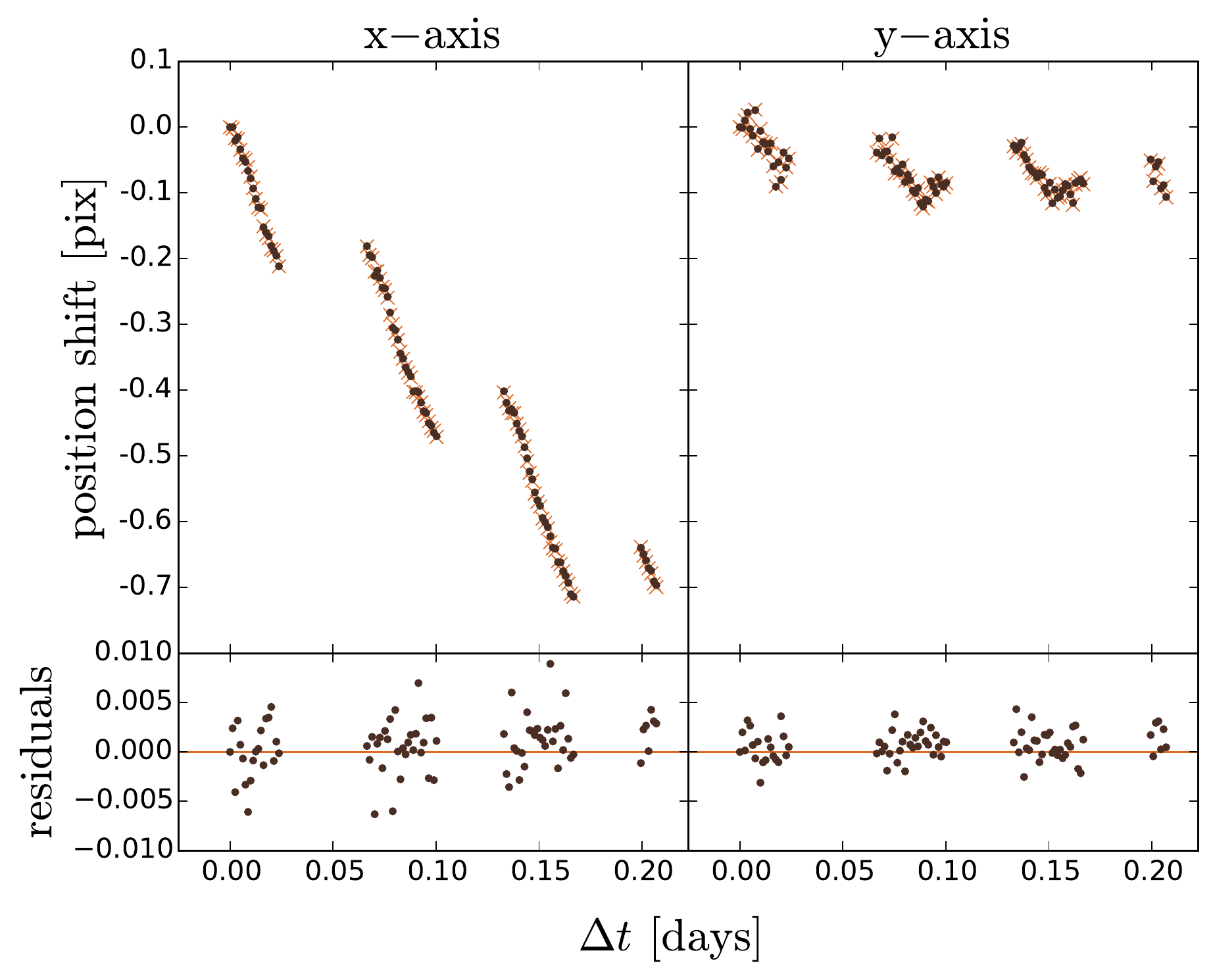}
\caption{Retrieved position shifts (dots) compared to the input values (crosses) and residuals for the two axes.}
\label{fig:position}
\end{figure}

\begin{figure}
\centering
\includegraphics[width=\columnwidth]{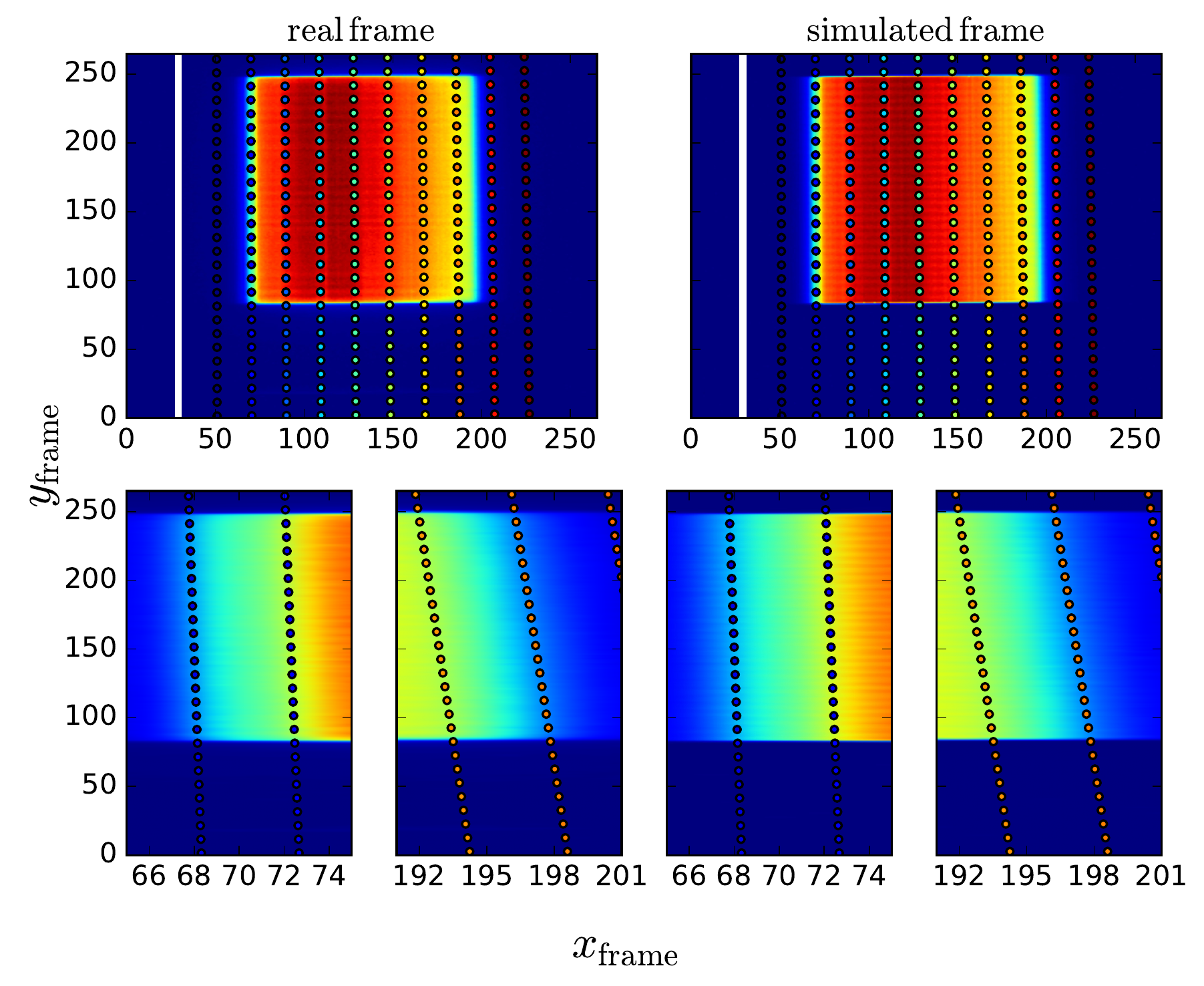}
\caption{Top: The wavelength-dependent photon positions (coloured points) as the star moves along its scanning trajectory (white line). Bottom: Left and right edges of the spectra where we can verify the similar to the real frame behaviour of the simulated frame.}
\label{fig:structure}
\end{figure}

At first, we apply all the reduction steps in the same way as implemented for a real data set, including: bias drifts correction, zero-read subtraction, non linearity correction, dark current subtraction, gain variations calibration, sky background subtraction and bad pixels/cosmic rays correction.

Then, the frames are calibrated for position shifts. For the (horizontal) x-shifts, we are making use of the normalised sum across the columns of the frames, while for (vertical) y-shifts we determine the position of the spectrum on the first non-destructive read. The input values that we use match the shifts presented in the real data set and, as it can be seen in Figure \ref{fig:position}, we are able to retrieve them very accurately. This result ensures that all the effects caused by the position-dependent systematics are properly simulated by \wfcsim.

To extract the 1D spectra from the spatially scanned spectra we first calculate the physical position of the star on the full-array detector. We then shift the position of the photons during the scan, and extract the flux for each wavelength bin from an aperture of quadrangular shape \citep{Tsiaras2015}. In Figure \ref{fig:structure} we can see the wavelength-dependent photon positions as calculated for a real and a simulated frame. From the similar structure we can conclude that the field dependent spectrum trace and wavelength solution are correctly implemented during the scanning process. 

\begin{figure}
\centering
\includegraphics[width=\columnwidth]{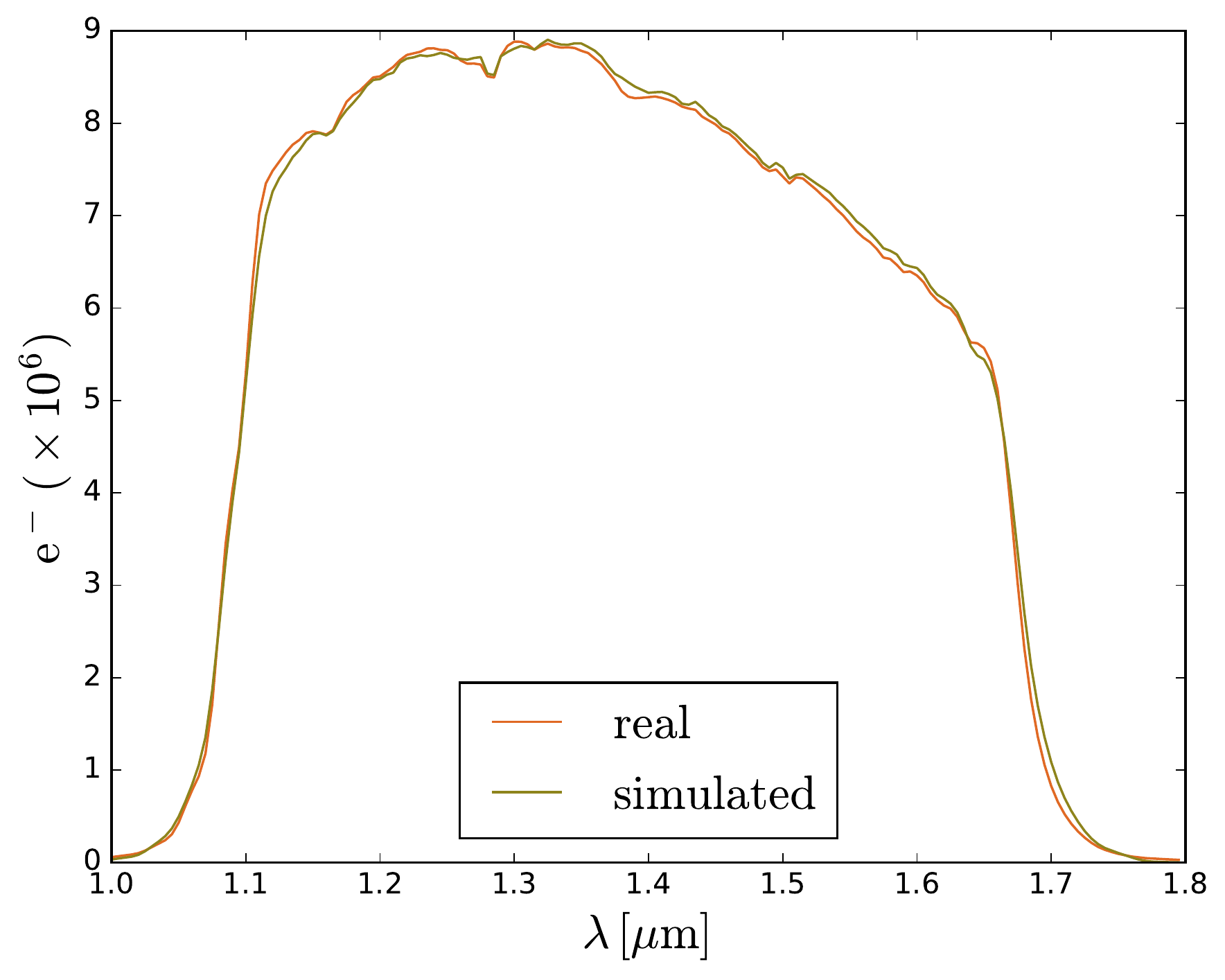}
\caption{1D spectra extracted from a real and a simulated spatially scanned frame, compared with the sensitivity curve of the \grisml\ grism. Although the stellar spectrum used is the closest to \system, we can see that the real spectrum is smoother, probably due to a lower number of absorption lines.}
\label{fig:1dspec}
\end{figure}

We, can therefore compare the 1D spectra, as extracted from a real and a simulated frame (Figure \ref{fig:1dspec}). The two spectra have the same shape, consistent with each other and the sensitivity curve of the \grisml\ grism. However, the real spectrum is smoother, probably due to a lower number of absorption lines.

\subsection{Fitting the white and spectral light-curves} \label{subsub:fitting}

The last step is to fit the white-light curve for the mid-transit point and the `ramp' coefficients and then use these values to fit the $R_\mathrm{p}/R_*$ for the different wavelength channels. We follow the same two approaches as we did for the real data set, i.e fitting the instrumental systematics function, Equation \ref{rampfunction}, and the transit model both separately and simultaneously, and keeping inclination, $i$, and $a/R_*$ ratio fixed. The results we obtain (Table \ref{tab:parameters}) are consistent with each other and in good agreement with the inputs. Also the magnitude of the errors are as expected for the signal-to-noise ratio of this particular target observed by \wfc.

\begin{figure}
    \centering
    \includegraphics[width=\columnwidth]{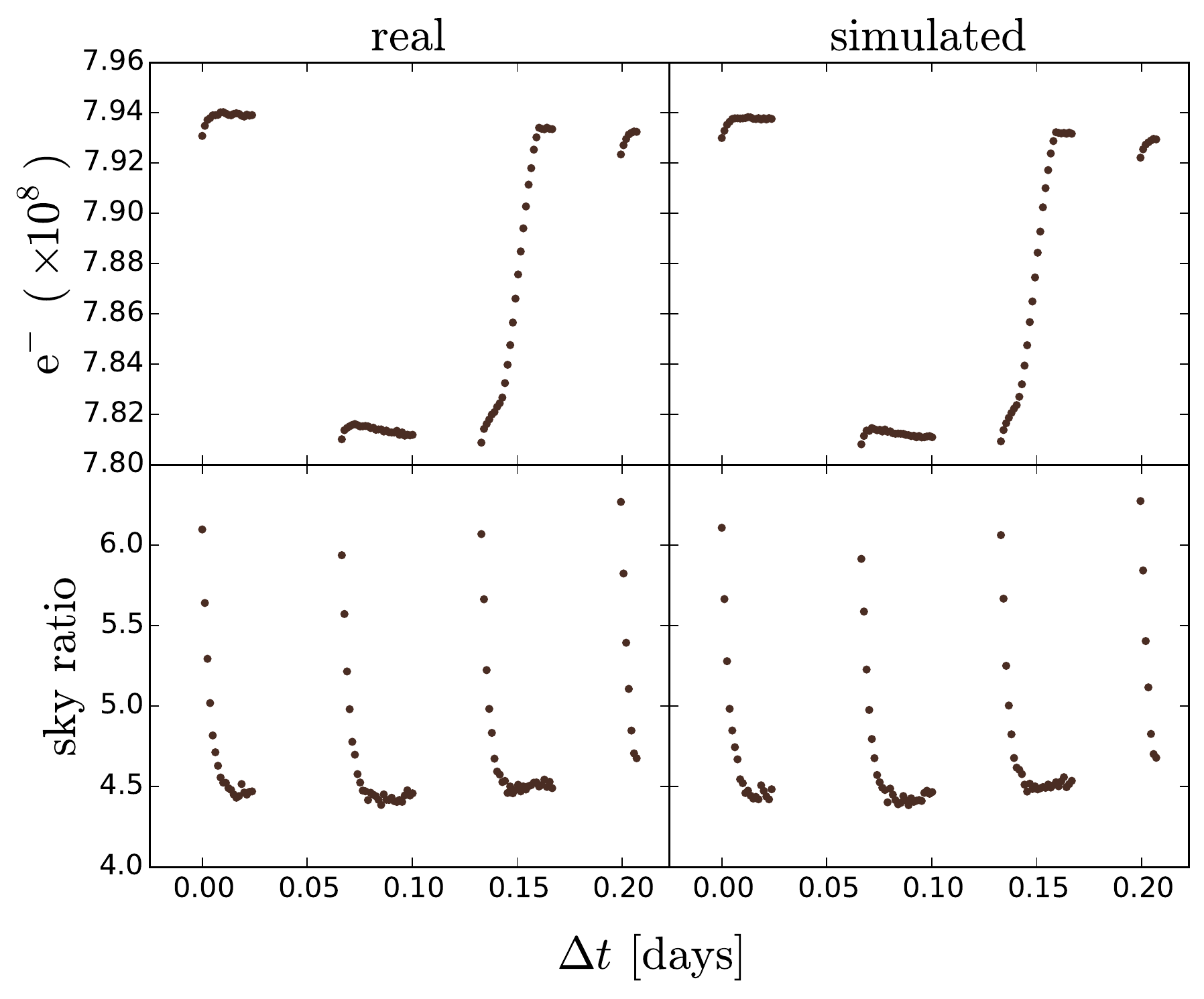}
    \caption{Top: White light-curve, as extracted from the real and the simulated data sets. Bottom: The ratio between the images and the master-sky frame for the real and the simulated data sets.}
    \label{fig:comp_lightcurve}
\end{figure}

\begin{figure}
    \centering
    \includegraphics[width=\columnwidth]{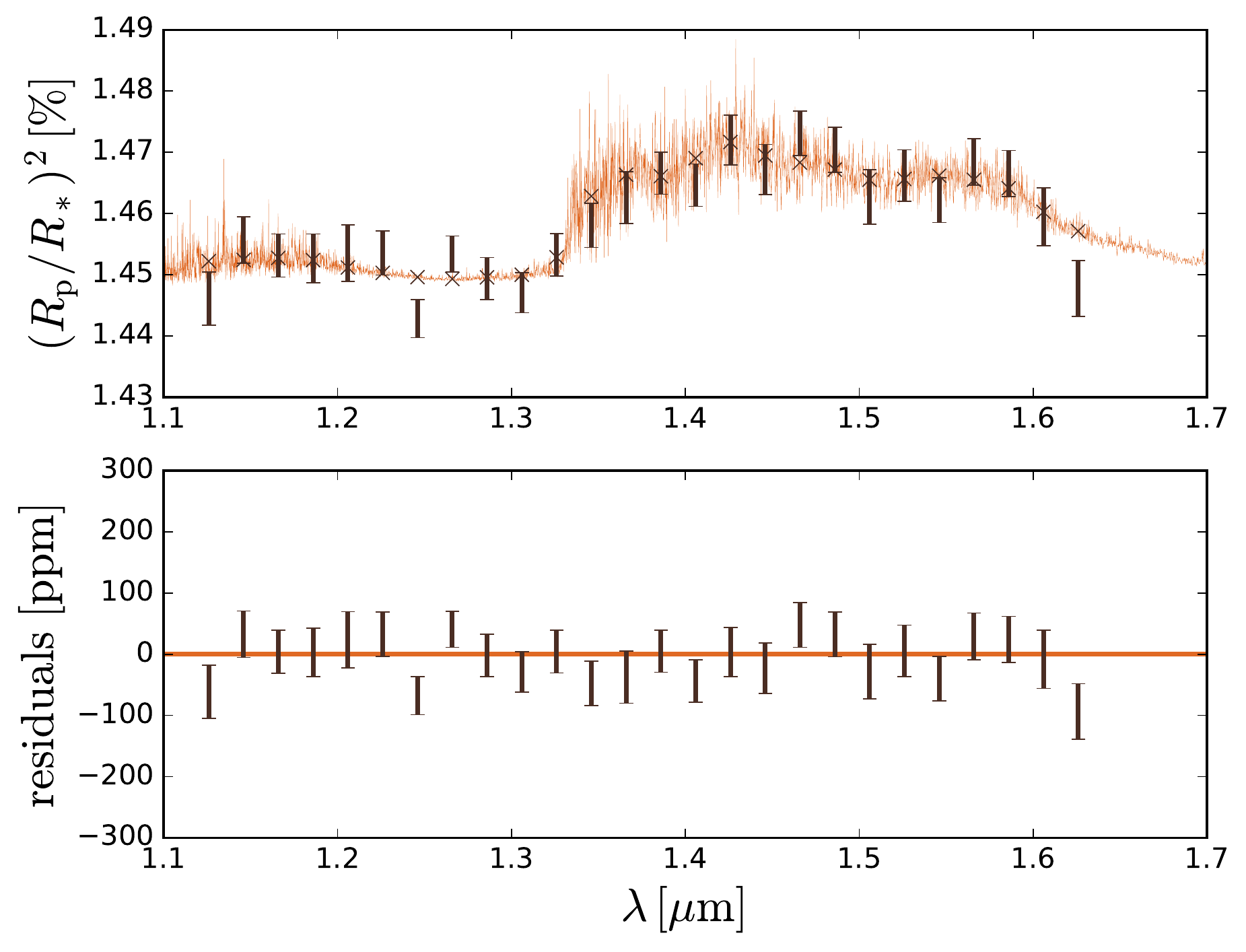}
    \caption{High resolution input spectrum of \planet\ and the extracted data from the simulation. In the bottom panel we can see the residuals, which have a rms of 38\,ppm.}
    \label{fig:comp_spectrum}
\end{figure}

We then fit the spectral light-curves leaving as free parameters only the normalisation factor and the $R_\mathrm{p}/R_*$ ratio. As explained in \citet{Tsiaras2015}, the spectral light-curves include an additional kind of systematics which originates from the low resolution of the spectrum (under-sampling) coupled with the horizontal shifts \citep{Deming2013}. Since the simulated data set includes these shifts we have to take these systematics into account. The final results can be seen in Figure \ref{fig:comp_spectrum}. The spectrum extracted from the simulated data has a mean uncertainty of 38\,ppm, slightly better than the real one (40\,ppm). The rms of the residuals between the input and the output spectra is also 38\,ppm and it is consistent throughout 10 different simulations. Based on these results we can conclude that simulations created by \wfcsim\ can reproduce existing data sets in a realistic way.

\section{Conclusion}

We implemented the \wfcsim\ algorithm as a Python package capable of simulating Hubble Space Telescope (\hst) Wide Field Camera 3 (\wfc) grism spectroscopy, including sources of noise and systematics. From all our diagnostics we conclude that given the configuration of a real data set, \wfcsim\ generates a simulated data set with the same behaviour and noise level as the real case. Therefore, it is a powerful tool with which we can validate the stability of existing data sets with certain configuration but also predict the behaviour of future observations in order to decide the best observing strategy for a particular target. 

\newpage

\wfcsim\ can then be used to test how different analysis methods work in different scenarios in order to validate data analysis pipelines. We demonstrate in this initial investigation that the method of \citet{Tsiaras2015} can accurately recover the independent planet signal for that data set.

Whilst written primarily for \hst\ \wfc, \wfcsim\ was designed to be easily adaptable to new instruments and systematics, making it a powerful and versatile tool to assess analysis techniques on both current and proposed instrument designs on different telescopes.

\acknowledgments
\section*{Acknowledgments}
We would like to thank Giovanna Tinetti, Ingo Waldmann, Marco Rocchetto and Giuseppe Morello for useful discussions about this project. This work has been supported by a UCL IMPACT studentship, STFC (ST/P002153/1) and ERC project 617119 (ExoLights). 

\newpage

\appendix

\vspace{1cm}

\section{Additional FITS header keywords used by Wayne}
\vspace{-0.5cm}
\begin{table}[h!]
\footnotesize
\center
\label{tab:fits-headers}
\caption{Keywords added to the science header of the output FITS files that are not in normal \stsci\ files. Values for the 0001\_raw.fits files of the \planet\ case study simulation are included as an example.}
	\begin{tabular}{lll}
					& Value in	0001\_raw.fits	&										\\ 
	Header Keyword	& from case study		&  Description								\\
	\hline
	SIM				& T					& used to clearly identify if the file is simulated		\\ 
	SIM-VER			& 1.0.0.dev1			& Wayne version used						\\ 
	SIM-TIME			& 36.37				& time taken to generate exposure (s)			\\ 
	X-REF 			& 404.4970			& x reference position of star on frame (full frame)	\\ 
	Y-REF			& 457.4293			& y reference position of star on frame (full frame)	\\ 
	SAMPRATE		& 0.02				& subsample time (s)						\\ 
	NSE-MEAN		& F					& mean of extra gaussian noise					\\ 
	NSE-STD			& F					& standard deviation of extra gaussian noise		\\ 
	ADD-DRK			& T					& dark file added (T/F)						\\ 
	ADD-FLAT		& T					& flat field added (T/F)						\\ 
	ADD-GAIN		& T					& gain variations added (T/F)					\\ 
	ADD-NLIN		& T					& non-linearity effects added (T/F)				\\ 
	STAR-NSE		& T					& stellar noise Added (T/F)					\\ 
	CSMCRATE		& 11					& rate of cosmic rays (hits per s)				\\ 
	SKY-LVL			& 4.63477				& sky background level (master sky per s)			\\ 
	VSTTREND		& 0.9989				& visit trend scale factor						\\ 
	CLIPVALS		& T					& pixels clipped to detector range (T/F)			\\ 
	RANDSEED		& 0					& random seed used for the visit				\\ 
	V-PY				& 2.7.11 final			& Python version used						\\ 
	V-NP				& 1.11.0				& NumPy version used						\\ 
	V-SP				& 0.16.1				& SciPy version used						\\ 
	V-AP				& 1.1.2				& Astropy version used						\\ 
	V-PD				& 0.18.1				& Pandas version used						\\ 
	MID-TRAN		& 2456196.28835		& Time of mid transit (HJD)					\\ 
	PERIOD			& 3.52474859			& Orbital period (days)						\\ 
	SMA				& 8.8587				& Semi-major axis (a/R$_s$)					\\ 
	INC				& 86.71				& Orbital inclination (deg)						\\ 
	ECC				& 0.00				& Orbital eccentricity							\\ 
	PERI				& nan				& Argument of periastron (deg)					\\ 
	LD1				& 0.608402			& Non-linear limb darkening coefficient 1			\\ 
	LD2				& -0.206180			& Non-linear limb darkening coefficient 2			\\ 
	LD3				& 0.262286			& Non-linear limb darkening coefficient 3			\\ 
	LD4				& -0.133088			& Non-linear limb darkening coefficient 4			\\ 
	\end{tabular}
\end{table}

\newpage

\section{\wfc\ correction files used}
\vspace{-0.5cm}
\label{sec:correction-files}

\begin{table}[h!]
	\footnotesize
	\center
	\label{tab:calib-files}
	\caption{The file names and download locations of files that were mentioned in the text.}
	\begin{tabular}{l l}
	File						& File name					\\
	\hline
	\grisml\ Sensitivity\,$^a$		& WFC3.IR.G141.1st.sens.2.fits	\\
	\grisml\ Master Sky\,$^a$		& WFC3.IR.G141.sky.V1.0.fits		\\
	\grisml\ Flat-field Cube\,$^a$	& WFC3.IR.G141.flat.2.fits		\\
	\grisms\ Sensitivity\,$^b$		& WFC3.IR.G102.1st.sens.2.fits	\\
	\grisms\ Master Sky\,$^b$		& WFC3.IR.G102.sky.V1.0.fits		\\
	\grisms\ Flat-field Cube\,$^b$	& WFC3.IR.G102.flat.2.fits		\\
	Superdarks\,$^c$			& u4819490i\_drk.fits, u481949ri\_drk.fits, xag19296i\_drk.fits, xcc20398i\_drk.fits	\\
							& u4819491i\_drk.fits, u481949ti\_drk.fits, xag19297i\_drk.fits, xcc20399i\_drk.fits	\\
							& u4819493i\_drk.fits, u4819501i\_drk.fits, xag19298i\_drk.fits, xcc2039ai\_drk.fits	\\
							& u4819494i\_drk.fits, x5g1509ki\_drk.fits, xag19299i\_drk.fits, xcc2039bi\_drk.fits	\\
							& u481949ji\_drk.fits, xag19292i\_drk.fits, xag1929ai\_drk.fits, xcc2039ci\_drk.fits	\\
							& u481949ki\_drk.fits	, xag19293i\_drk.fits, xcc20394i\_drk.fits, xcc2039di\_drk.fits	\\
							& u481949li\_drk.fits, xag19294i\_drk.fits, xcc20396i\_drk.fits, u481949mi\_drk.fits	\\
							& xag19295i\_drk.fits, xcc20397i\_drk.fits	\\
	Quantum Efficiency			& wfc3\_ir\_qe\_003\_syn.fits		\\
	Non-Linearity\,$^c$			& u1k1727mi\_lin.fits				\\
	Gain Variations\,$^c$		& u4m1335mi\_pfl.fits			\\
	Initial Bias	\,$^d$			& wfc3\_ir\_initial\_bias\_256.fits	\\
	\multicolumn{2}{l}{$^a$ \url{http://www.stsci.edu/hst/wfc3/analysis/grism_obs/calibrations/wfc3_g141.html}} \\
	\multicolumn{2}{l}{$^b$ \url{http://www.stsci.edu/hst/wfc3/analysis/grism_obs/calibrations/wfc3_g102.html}} \\
	\multicolumn{2}{l}{$^c$ \url{http://www.stsci.edu/hst/observatory/crds/SIfileInfo/WFC3/}} \\
	\multicolumn{2}{l}{$^d$ \url{http://www.ucl.ac.uk/exoplanets/wayne/}}
	\end{tabular}
\end{table}

\section{Grism Calibration Coefficients}
\vspace{-0.5cm}

\begin{table}[h!]
\footnotesize
\centering
\caption{Coefficients of the field dependent trace for the \grisms\ and \grisml\ grisms as given by \citet{Kuntschner2009G102, Kuntschner2009G141}}
\label{tab:grism-trace-coeff}
\begin{tabular}{l | lllllllll}
& $a_0$ & $a_1 (X)$ & $a_2 (Y)$ & $a_3 (X*Y)$ & $a_4 (X^2)$ & $a_5 (Y^2)$ \\
\hline
\grisms\ $A_c$ & -3.55018E-1 &  3.28722E-5 & -1.44571E-3  & & & \\
error      &  7.40459E-2 &  4.4456E-6  &  3.653212E-6 & & & \\
\grisms\ $A_m$ & 1.42852E-2 & -7.20713E-6 & -2.42542E-6 & 1.18294E-9 & 1.19634E-8 & 6.17274E-10 \\
error      & 3.86038E-4 &  4.21303E-7 &  3.42753E-7 & 4.26462E-10 & 3.51491E-10 & 3.02759E-10 \\
\grisml\ $A_c$ & 1.96882 & 9.09159E-5 & -1.93260E-3 & & & \\
error      & 8.09111E-2 & 3.57223E-6 & 3.12042E-6 & & & \\
\grisml\ $A_m$ & 1.04275E-2 & -7.96978E-6 & -2.49607E-6 & 1.45963E-9 & 1.39757E-8 & 4.8494E-10 \\
error      & 5.94731E-4 & 4.34517E-7 & 3.57986E-7 & 3.87141E-10 & 3.29421E-10 & 3.08712E-10 \\
\end{tabular}
\end{table}

\begin{table}[h!]
\footnotesize
\centering
\caption{Coefficients of the field dependent wavelength solution for the \grisms\ and \grisml\ grisms as given by \citet{Kuntschner2009G102, Kuntschner2009G141}}
\label{tab:grism-wl-sol-coeff}
\begin{tabular}{l | lllllllll}
& $b_0$ & $b_1 (X)$ & $b_2 (Y)$ & $b_3 (X*Y)$ & $b_4 (X^2)$ & $b_5 (Y^2)$ \\
\hline
\grisms\ $B_c$ &  6.38738E3 & 4.55507E-2 & 0 & & & \\
error      & 3.17621 & 3.19685E-3 & -  & & & \\
\grisms\ $B_m$ & 2.35716E1 & 3.60396E-4 & 1.58739E-3 & -4.25234E-7 & -6.53726E-8 & 0 \\
error      & 2.33411E-2 & 1.49194E-4 & 1.05015E-4 & 1.80775E-7 & 9.35939E-8 & - \\
\grisml\ $B_c$ &  8.95431E3 & 9.35925E-2 & 0 & & & \\
error      &  8.14876 & 1.09748E-2 & - & & & \\
\grisml\ $B_m$ &  4.51423E1 &  3.17239E-4 & 2.17055E-3 & -7.42504E-7 & 3.48639E-7 & 3.09213E-7 \\
error      &  6.26774E-2 & 3.98039E-4 & 2.3185E-4 &   4.45730E-7 & 3.20519E-7 & 2.16386E-7 \\
\end{tabular}
\end{table}

\begin{table}[h!]
\footnotesize
\centering
\caption{PSF Coefficients}
\label{tab:psf-coeff}
\begin{tabular}{c | c c c c}
& $c_3 (\lambda^3)$  & $c_2 (\lambda^2)$ & $c_1 (\lambda)$ & $c_0$   \\
\hline
$\alpha$ & -0.2506 & 0.8332 & -0.8055 & 0.3990  \\
$\sigma_1$ & 0.6925 & -2.1043 & 2.223 & -0.2969 \\
$\sigma_2$ & 2.904 & -8.819 & 8.960 & 2.255
\end{tabular}
\end{table}

\newpage

\bibliographystyle{apj}
\bibliography{wayne,wayne-man} 

\end{document}